\documentclass[12pt,a4paper,dvips]{article}
\usepackage{cite,mcite}
\usepackage{graphicx}
\graphicspath{{/afs/cern.ch/l3/paper/example/figs/}}
\newlength{\capindent}
\newlength{\capwidth}
\newlength{\figwidth}
\newcommand{\icaption}[2][!*!,!]{\hspace*{\capindent}%
  \begin{minipage}{\capwidth}
    \ifthenelse{\equal{#1}{!*!,!}}%
      {\caption{#2}}%
      {\caption[#1]{#2}}
  \end{minipage}}
\def\a34{\cos\alpha_{34}}

\def\ra{\rightarrow}

\def\ee{\mathrm{e^{+}e^{-}}}
\def\AFB{\mathrm{\rm A_{FB}\;}}

\def\COST{\mathrm{\rm \cos\theta\;}}

\def\EE{\mathrm{\rm \;e^+e^-\;}}

\def\GG{\mathrm{\rm \;\gamma\gamma\;}}
\def\GGG{\mathrm{\rm \;\gamma\gamma(\gamma)\;}}
\def\MM{\mathrm{\;\mu^+\mu^-\;}}
\def\TT{\mathrm{\;\tau^+\tau^-\;}}
\def\l{\mathrm{\lambda}}
\def\L{\mathrm{\Lambda}}
\def\SP{\mathrm{s^{\prime}}}

\def\qq{\mathrm{q\bar{q}}}

\def\rs{\sqrt{s}}

\def\xi{x_{i}}

\def\TeV{\mathrm{TeV}}
\def\LL{\mathrm{LL}}
\def\RR{\mathrm{RR}}
\def\RL{\mathrm{RL}}
\def\LR{\mathrm{LR}}
\def\O{\mathrm{\not\! O}}

\newcommand {\Be}{\begin{equation}}
\newcommand {\Ee}{\end{equation}}

\newcommand {\eqref}[1]{equation~(\ref{#1})}
\newcommand {\Eqref}[1]{Equation~(\ref{#1})}

\newcommand {\Figref}[1]{Figure~\ref{fig:#1}}

\newcommand {\Tabref}[1]{Table~\ref{tab:#1}}

\renewcommand{\thefootnote}{\fnsymbol{footnote}}

\begin{document}

\begin{titlepage}
\begin{flushright} 
hep-ex/0103039 \\
%%hep-ph/0002172 \\
%%ETHZ-IPP PR-2000-02 \\
% hep-ex/9806027 \\
March 22, 2001
\end{flushright}
%\voffset=-20pt
%\begin{figure}[h]
%\includegraphics*[scale=1.,angle=0,bb=222 438 402 552 ]
%{ethlogo.ps}
%\end{figure}
%\begin{picture}(100,-100)(10,20)
%\put (350.,130.){\bf ETHZ-IPP  PR-2000-02}
%\end{picture}
%\begin{picture}(100,-100)(10,20)
%\put (245.,112.){\bf February 16, 2000}
%\end{picture}

\vspace*{2.0cm}

\begin{center} {\Large \bf
                         Two--fermion and Two--photon Final States \\
\vspace*{0.15cm}
                         at LEP2 and Search for Extra Dimensions}

\vspace*{2.5cm}
  {\Large
  Dimitri Bourilkov\footnote{\tt e-mail: Dimitri.Bourilkov@cern.ch,$\ \ \ $
                                 Web: http://cern.ch/bourilkov/}
  }

\vspace*{1.0cm}
  Institute for Particle Physics (IPP), ETH Z\"urich, \\
  CH-8093 Z\"urich, Switzerland\\
\vspace*{1.5cm}
{Invited LEP2 review talk\\
{\em Les Rencontres de Physique de la Vall\'ee d'Aoste}\\
La Thuile, Aosta Valley, Italy, 4-10 March 2001}
\end{center}

%\date{June 4, 1999}
%\maketitle
%       
%
% The abstract
%
\vspace*{1.2cm}
\begin{abstract}
For the first time the experiments ALEPH, DELPHI, L3 and OPAL
have presented {\it preliminary} results for fermion-pair and
photon-pair production in $\ee$ collisions on the full LEP2
data set. The details of the experimental measurements and
results from their LEP--wide combination are presented.
No statistically significant deviations from the Standard
Model expectations are observed and lower limits,
some obtained in dedicated analyses,
for new physics phenomena at 95~\% confidence level are derived.
The scales of contact interactions are constrained to lie
above 10--20 TeV, depending on the helicity structure.
The Standard Model has thus been tested at LEP2 down to
distances $\rm 10^{-19} - 10^{-20}$~m.
This has many implications, one example being the interpretation
of the new result on the anomalous magnetic moment of the muon
as coming from muon substructure.
For $\GG$ final states the QED cut-offs are
$\L_+ > 0.44\ \TeV$ and $\L_- > 0.37\ \TeV$.
In a combined analysis, using $\ee$ and $\GG$ final states,
the most stringent lower limits to date,
\mbox{$M_s^+ > 1.13\;\TeV\ (\l = +1)$} and
\mbox{$M_s^- > 1.39\;\TeV\ (\l = -1)$},
on the low gravity effective Planck scale are set.
Constrains on the scales of string models like {\it TeV strings} and
{\it D-branes} are derived. In the last case the lower limit
is $\rm 1.5 - 4\ \TeV$, depending on the coupling strength. 
\end{abstract}

\vspace*{1.5cm}

\end{titlepage}

\renewcommand{\thefootnote}{\arabic{footnote}}
\setcounter{footnote}{0}
%
%%%%%%%%%%%%%%%%%%%%%%%%%%%%%%%%%%%%%%%%%%%%%%%%%%%%%%%%%%%%%%%%%%%%%%%%%%%%%%%
% Introduction
%%%%%%%%%%%%%%%%%%%%%%%%%%%%%%%%%%%%%%%%%%%%%%%%%%%%%%%%%%%%%%%%%%%%%%%%%%%%%%%
%
\section{Introduction}

\subsection*{Data}

\hspace*{0.7cm}

The four LEP collaborations ALEPH, DELPHI, L3 and OPAL have
taken data at energies above the Z resonance from 1995,
starting at 130 GeV, until November 2000, culminating
at energies close to 209 GeV.
For the first time we have results on the full LEP2 data set,
giving a flavor of the expected final experimental precision
and sensitivity to physics beyond the Standard Model (SM).
With minor exceptions, the data up to 189 GeV centre-of-mass energy
are final, and the data collected in 1999 and 2000 at energies between
192--209 GeV {\em preliminary}. The luminosity accumulated above
the Z pole in a single experiment is summarized in~\Tabref{lumi}.

\begin{table}[h]
\renewcommand{\arraystretch}{1.1}
\caption{Data collected in a single experiment above the Z resonance.}
\label{tab:lumi}
 \begin{center}
  \begin{tabular}{|c|c|c|c|c|}
\hline
  Year      & $\sqrt{s}$ & $\mathcal{L}$/exp. & $\sigma_{\qq}$ - stat. err. [\%]/exp.& Mode of\\
            &  GeV       &    pb$^{-1}$  & total / HE    & operation \\
\hline
1995 \& 97  &   130      &         6     & 2.3/5.2    & Above \\
1995 \& 97  &   136      &         6     & 2.6/6.0    & $M_Z$ \\
\hline                   
1996        &   161      &         10    & 2.7/5.9    & W pairs \\
1996        &   172      &         10    & 3.2/7.3    & W mass  \\
\hline                   
1997        &   183      &         55    & 1.4/3.3    & Z pairs \\
\hline                   
1998        &   189      &     177  &     0.8/1.8  & High  \\
1999        &   192-202  &  233  &  0.7/1.6  & precision  \\
\hline                   
2000        &   202-209  &    220  &    0.7/1.6  & Higgs   \\
\hline
\hline
1995-2000   &   130-209  &     720  &     0.4/0.8  & LEP2    \\
\hline
  \end{tabular}
 \end{center}
\end{table}

In the same table the statistical error on the hadronic cross section is
given in \%. The combined error for the total sample is $\sim 0.4$\%,
and for the most interesting non-radiative (or genuine high energy)
sample $\sim 0.8$\%. If we combine the measurements of the four
experiments the statistical error is reduced by a factor of two.
Clearly LEP2 has reached the phase of high precision, putting
additional requirements on the careful study of systematic effects
and on matchingly accurate theory predictions.

\subsection*{Fermion-pair Production}

\begin{figure}[h]
\begin{center}
\resizebox{1.0\textwidth}{4.6cm}{\includegraphics*[1cm,22.4cm][14cm,26cm]{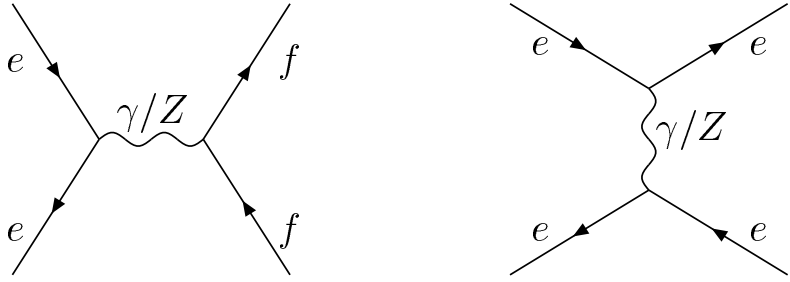}}
\end{center}
  \caption{\em Standard Model Feynman diagrams for difermion final states.}
  \label{fig:figure1}
\end{figure}

\hspace*{0.7cm}
In the Standard Model the production of a fermion-pair in $\ee$ collisions
is described by the exchange of $\gamma$ or Z in the $s$-channel, and if the
final state is identical to the initial one, also in the $t$-channel
({\it cf.}~\Figref{figure1}).

The differential cross section has the form:
\Be
\frac{d \sigma}{d \Omega} = |\gamma_s+Z_s+elec*(\gamma_t+Z_t) +  New\;Physics\;?!|^2
\label{eq1}
\Ee
where the contributing amplitudes are denoted symbolically by $\gamma_s$, $Z_s$,
$\gamma_t$ and $Z_t$.

\begin{figure}[h]
\begin{center}
\vspace{-1.0cm}
\resizebox{0.8\textwidth}{11cm}{\includegraphics{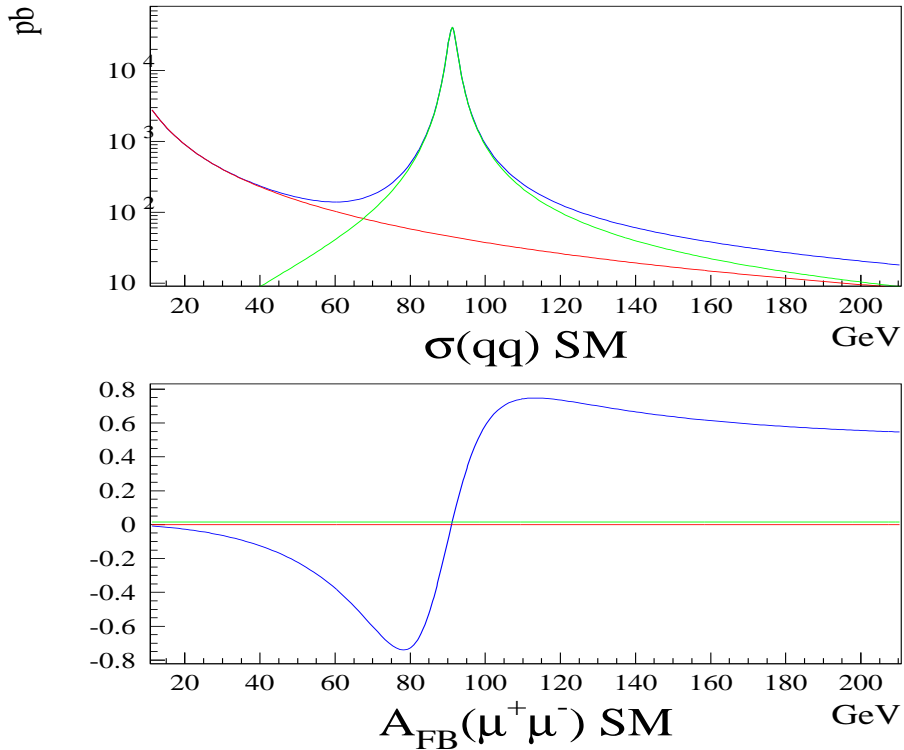}}
\end{center}
  \caption{\em Hadron cross section and muon forward-backward asymmetry
           in $\ee$ collisions (red - photon exchange, green - Z exchange,
           blue - total).}
  \label{fig:figure2}
\end{figure}

As an illustration the hadron cross is given in~\Figref{figure2} as a function
of the centre-of-mass energy. The contributions from photon or Z exchange
are shown separately together with the total cross section, which includes
also interference effects. We can observe that at 200 GeV the contribution
from photon exchange has a size similar to the one coming from Z exchange.
For leptonic final states, due to the higher electric charges,
the photon exchange dominates.
In the same figure the forward-backward asymmetry
for muons is shown. Here the contributions from photon or Z exchange
show up as straight lines close to zero, and the large positive asymmetry
above the Z pole is totally dominated by the interference term.

The interest in studying fermion-pair final states at LEP2 is driven by
the fact that many types of new physics scenarios can contribute to these
processes, as shown by the last amplitude in~\Eqref{eq1}. For this to 
happen, the couplings to the initial and final states should be different
from zero. In this case, even if the Standard Model extension operates
at an energy scale much higher than the accessible centre-of-mass energy
for a direct observation, it can still give measurable effects by modifying
the differential cross section through interference with the SM amplitudes.

\subsection*{Contact Interactions}

\hspace*{0.7cm}
Contact interactions offer a general framework for describing 
a new interaction with typical energy scale  $\L \gg \rs$.
The presence of operators with canonical dimension
 $N > 4$ in the Lagrangian gives rise to effects  $\sim 1/M^{N-4}$.
Such interactions can occur for instance, if the SM particles are composite, or
when new heavy particles are exchanged.

\begin{figure}[h]
\begin{center}
\resizebox{1.0\textwidth}{4.6cm}{\includegraphics*[1cm,22.4cm][14cm,26cm]{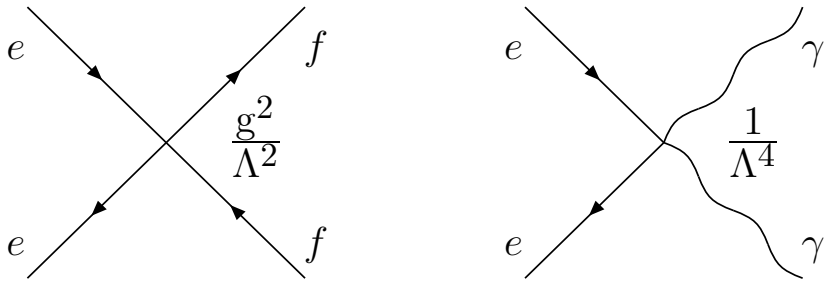}}
\end{center}
  \caption{\em Contact interaction diagrams for difermion or diphoton final states.}
  \label{fig:figure3}
\end{figure}

For fermion- or photon-pair production the contact interactions are described by
the diagrams shown in~\Figref{figure3}.
In the fermion case, the lowest order flavor-diagonal and helicity-conserving
operators have dimension six~\cite{PeskinCI}.
For photons the lowest order operators have dimension eight, leading to
suppression of the interference terms as the inverse forth power of the
relevant energy scale.
%{\Large 
%$${\mathcal A}^{ef}_{LR(RL)}(s) = ({\mathcal A}^{SM}_{LR(RL)}+\eta_{LR(RL)}\frac{s}{\alpha}\frac{1}{\L^2})\frac{t}{s}$$
%$${\mathcal A}^{ef}_{LL(RR)}(s) = ({\mathcal A}^{SM}_{LL(RR)}+\eta_{LL(RR)}\frac{s}{\alpha}\frac{1}{\L^2})\frac{u}{s}$$
%}

The differential cross section takes the form
\Be
\frac{d \sigma}{d \Omega} = SM(s,t)+\varepsilon\cdot C_{Int}(s,t)+\varepsilon^2\cdot C_{NewPh}(s,t)
\Ee
where the first term is the Standard Model contribution, the second comes from
interference between the SM and the contact interaction, and the third is the
pure contact interaction effect.
The Mandelstam variables are denoted as $s$, $t$ and $u$.

Usually the coupling is fixed,
and the structure of the interaction is parametrized by coefficients
for the helicity amplitudes:
\begin{center}
\begin{tabular}{ll}
 g            &  coupling (by convention $\frac{g^2}{4\pi}=1$) \\
% $\L$  &  energy scale (or exchanged mass) \\
 $|\eta_{ij}|\leq 1$  & helicity amplitudes ($i,j = \rm L,R$)\\
 $\varepsilon$  & $\frac{g^2}{4\pi}\frac{sign(\eta)}{\L^2}$ for $f\bar f$; $\ \ \sim \frac{1}{\L^4}$ for $\GG$ 
\end{tabular}
\end{center}

Some often investigated models are summarized in~\Tabref{ci-models}. The models
in the second half of the table are parity conserving, and hence not
constrained by the very precise measurements of atomic parity violation at low
energies.
The results presented in this contribution cover the models in the table and,
as we will see in a moment, have connection to the recent searches for extra
spatial dimensions.
\begin{table}[h]
\renewcommand{\arraystretch}{1.20}
\caption{Contact interaction models.}
\label{tab:ci-models}
  \begin{center}
    \begin{tabular}{|c|cccc|cccc|}
\hline
~~Model~~&~~~LL~~&~~~RR~~&~~~LR~~&~~~RL~~&~~~VV~~
&~~~AA~~&~LL+RR~&~LR+RL~\\
        & \multicolumn{4}{c|}{ Non-parity conserving } & \multicolumn{4}{c|}{ Parity conserving } \\
  \hline \hline
$\eta_{\LL}$
   & $\pm$1& 0    &    0 &    0 &$\pm$1&$\pm$1&$\pm$1&  0   \\
$\eta_{\RR}$
   & 0     &$\pm$1&    0 &    0 &$\pm$1&$\pm$1&$\pm$1&  0    \\
$\eta_{\LR}$
   & 0     & 0    &$\pm$1&    0 &$\pm$1&$\mp$1& 0   &$\pm$1  \\
$\eta_{\RL}$
   & 0     & 0    &    0 &$\pm$1&$\pm$1&$\mp$1& 0   &$\pm$1 \\
\hline
    \end{tabular}
  \end{center}
\end{table}

\begin{figure}[!ht]
\begin{center}
\vspace{-1.0cm}
\resizebox{0.8\textwidth}{11cm}{\includegraphics{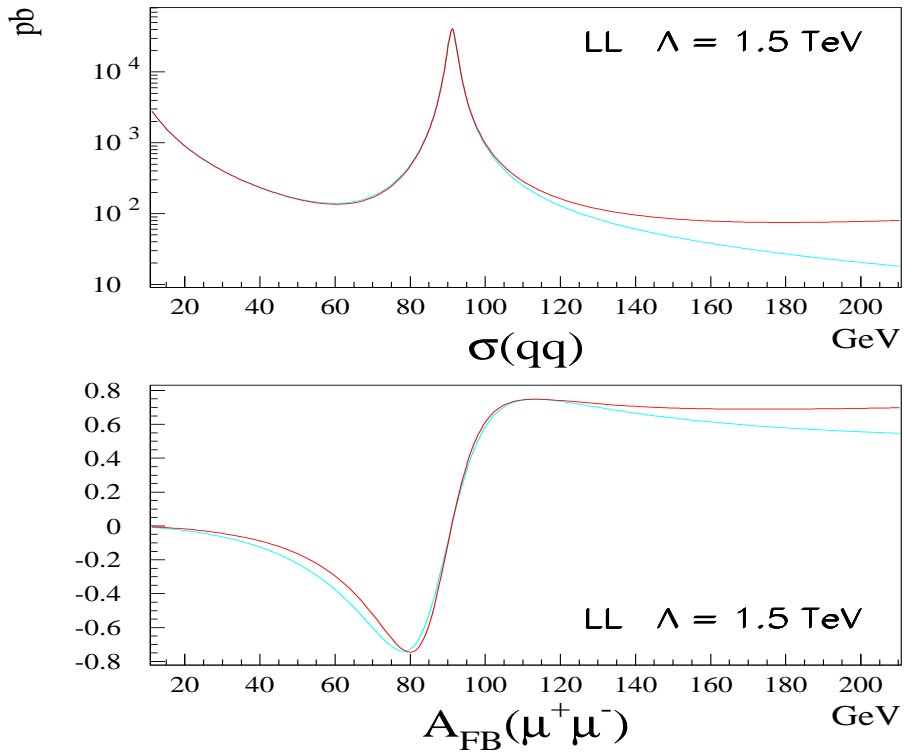}}
\end{center}
  \caption{\em Hadron cross section and muon forward-backward asymmetry
           in $\ee$ collisions in the presence of contact interactions
           (cyan - Standard Model, red - with contact interactions).}
  \label{fig:figure3ci}
\end{figure}

As an illustration the effects of contact interactions for one
particular model are shown in~\Figref{figure3ci}.

\subsection*{Low Scale Gravity}

\hspace*{0.7cm}
The fastest progress in physics comes when we deepen our understanding
of space-time.
The development of string theory points to the existence of up
to seven additional dimensions, which are compactified at
very small distances, initially estimated to be $\sim 10^{-32}$~m,
and hence far below the scales probed at high energy colliders.
Recently, a radical proposal~\cite{ADD,*ADD2,*ADD3} has been
put forward for the solution of the hierarchy problem, which brings
close the electroweak scale $m_{EW} \sim 1\; \TeV$ and the
Planck scale $M_{Pl} \sim \frac{1}{\sqrt{G_N}} \sim 10^{15}\; \TeV$.
In this framework the effective four-dimensional $M_{Pl}$ is
connected to a new $M_{Pl(4+n)}$ scale in a (4+n) dimensional
theory:
\Be
 M_{Pl}^2 \sim M_{Pl(4+n)}^{2+n} R^n
\Ee
where there are n extra compact spatial dimensions of radius
$\sim R$, which could be as large as 1~mm.
This can explain the observed weakness of gravity at large distances.
At the same time,
quantum gravity becomes strong at a scale M of the order of few~TeV
and could have observable signatures at present and future colliders.
The attractiveness of this proposal is enhanced by the plethora of expected
phenomenological consequences described  by just a few parameters.

\begin{figure}[h]
\begin{center}
\resizebox{0.80\textwidth}{8.0cm}{\includegraphics{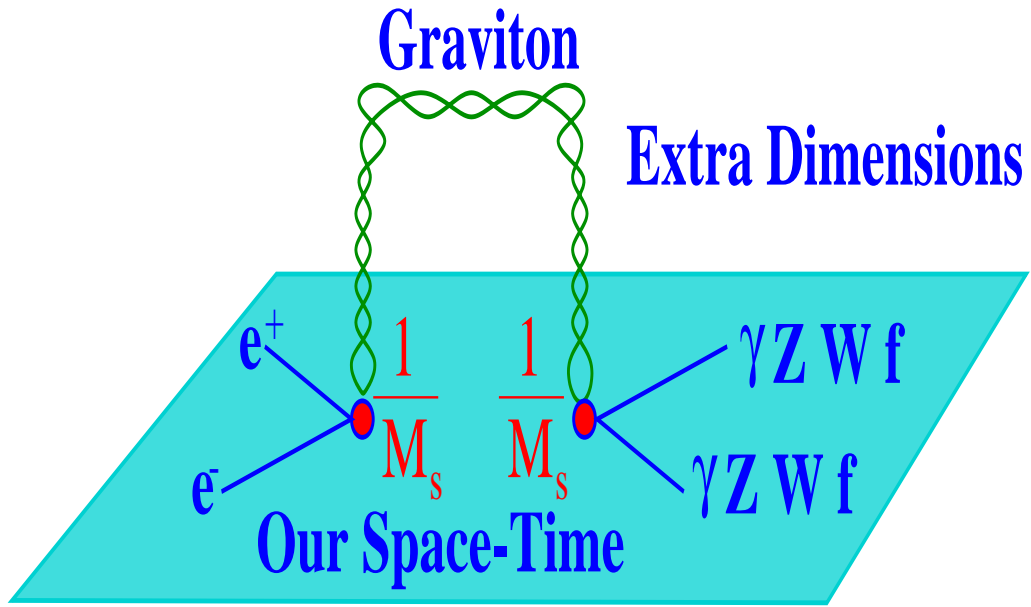}}
\end{center}
  \caption{\em Virtual graviton exchange diagrams for difermion or diboson final
            states.}
  \label{fig:figure4}
\end{figure}

In the production of fermion- or boson-pairs in $\ee$ collisions this class of
models can be manifested through virtual effects due to the exchange of
gravitons (Kaluza-Klein excitations), depicted in~\Figref{figure4}.
As discussed in~\cite{Hewett,Rizzo,Giudice,Lykken}, the exchange of spin-2
gravitons modifies in a unique way the differential cross sections for fermion
pairs, providing clear signatures. These models introduce an effective scale
(cut-off), denoted as $M_s$ in~\cite{Hewett,Rizzo}, as $\L_T$ in~\cite{Giudice}
and again as $M_s$ in~\cite{Lykken}.
The first two scales are connected by the relation $M_s=(2/\pi)^{1/4}\Lambda_T$,
which gives numerically \mbox{$\L_T = 1.1195\;M_s$}. They do not depend on the
number of extra dimensions. In the third case the scale exhibits such a
dependence; the relation to the other scales is given by 
\mbox{$M_s^{HLZ}|_{n=4} = \L_T$} for four extra dimensions.
As the majority of LEP experiments has decided to use the scale $M_s$
of~\cite{Hewett}, we will stick to this notation throughout this paper,
and convert the experimental results when necessary to facilitate the
comparison. The cut-off scale is supposed to be of the order of the
fundamental gravity scale M in 4+n dimensions.
The results are model-dependent, which is usually expressed by the
introduction of an additional parameter
\Be
\frac{\lambda}{M_s^4}.
\Ee
The value of $\lambda$ is not known exactly, the usual assumption is
$\lambda = \pm 1$ to allow for both constructive and destructive interference
effects.

\subsection*{String Models}

\hspace*{0.7cm}
The ideas outlined in the previous subsection have triggered new developments,
which propose concrete string realizations of the large extra dimensions
scenario. We will discuss two models of this type.

As noted in~\cite{Antoniadis,*Peskin}, the low-energy effective theory
that we are dealing with is derived from an underlying theory of quantum
gravity. And the only known consistent framework for the description of
quantum gravity is superstring theory. In string models the SM particles
are extended structures, represented by massless modes of open strings
on a set of branes. In this model massive string mode oscillations
(called {\em TeV strings}) give rise to dimension-8 operators which modify
the SM amplitudes and are expected to dominate over the effects coming from
Kaluza-Klein excitations. The model has two basic parameters: the dimensionless
Yang-Mills coupling constant $g_{YM}$ and the string scale - $M_S$~\footnote{
The effective gravity cut-off scale with subscript small s from the previous
subsection should not be confused with the string scale $M_S$,
studied here.}.
The relation to the fundamental gravitational scale M is depending on the
coupling
\Be
\frac{M}{M_S} = \left(\frac{1}{\pi}\right)^{1/8}\alpha^{-1/4};\ \ \ \ LSG\ scale\ M \sim 1.6 - 3.0\;M_S
\Ee
where $\alpha = g_{YM}^2/4\pi$, and the values 3.0 (1.6) correspond to
$\alpha = 1/137\ (\alpha_s(1\; \TeV))$ respectively.

In this model the tree-level open string four-point amplitudes are 
multiplied by a string form factor (first introduced by G.~Veneziano~\cite{Veneziano})
\Be
{\cal S}(s,t) = \frac{\Gamma(1-\frac{s}{M_S^2})\Gamma(1-\frac{t}{M_S^2})}{\Gamma(1-\frac{s}{M_S^2}-\frac{t}{M_S^2})}.
\Ee
For $\ee \ra \ee$ the differential cross section is modified:
\Be
\frac{d \sigma}{d \COST} = \left(\frac{d \sigma}{d \COST}\right)_{SM}\cdot |{\cal S}(s,t)|^2.
\Ee
For $\ee \ra \GG$ one can use Drell's parametrization (QED cut-off, {\it cf.} next subsection)
\Be
\L_+^{QED} = \left(\frac{12}{\pi^2}\right)^{1/4}\cdot M_S = 1.05\; M_S.
\Ee

In a new {\em D-brane model}~\cite{AntoniadisDBR,*AntoniadisReview},
the Standard Model matter
fields are identified with massless modes of open strings stretched between two
sets of branes - D3 and D7 branes (37 strings). In this scenario we can have
dimension-6 operators due to contact interactions induced by massive string
oscillators. These effects can be stronger than the ones coming from
Kaluza-Klein states or winding modes and will dominate the dimension-8
operators coming from TeV strings.

\subsection*{Photon-pair Production}

\begin{figure}[h]
\begin{center}
\resizebox{1.0\textwidth}{4.6cm}{\includegraphics*[1cm,22.4cm][14cm,26cm]{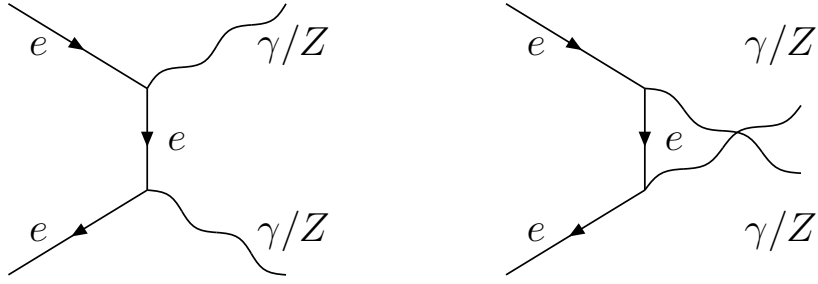}}
%\vspace{-1cm}
%\resizebox{\textwidth}{9cm}{\includegraphics[1cm,20cm][14cm,26cm]{../paper97/diag-ww.ps}}
\end{center}
  \caption{\em Feynman diagrams for diphoton (di-Z) final states.}
  \label{fig:figure5}
\end{figure}

\hspace*{0.7cm}
The production of photon pairs in $\ee$ collisions is described by the
$t$- and $u$-channel QED diagrams depicted in~\Figref{figure5}.

The differential cross section has the following simple form
\vspace{-0.2cm}
\Be
\frac{d \sigma}{d \Omega} = |e_t+e_u +  New\;Physics\;?!|^2
\Ee
\Be
\left(\frac{d \sigma}{d \Omega}\right)_{QED} = \frac{\alpha^2}{2s}\left[\frac{t}{u}+\frac{u}{t}\right] = \frac{\alpha^2}{s}\cdot\frac{1 + \cos^2\theta}{1 - \cos^2\theta}.
\Ee

The experiments have investigated many proposed deviations from QED:
\vspace{-0.2cm}
\begin{eqnarray}
\label{eq11}
\frac{d \sigma}{d \Omega} & = & \left(\frac{d \sigma}{d \Omega}\right)_{QED}\cdot \left(1 \pm \frac{1}{(\L_{\pm}^{QED})^4}\cdot \frac{s^2}{2}\sin^2\theta \right)  \\
\label{eq12}
\frac{d \sigma}{d \Omega} & = & \left(\frac{d \sigma}{d \Omega}\right)_{QED}\cdot \left(1 \mp \frac{\lambda}{\pi\alpha (M_s)^4}\cdot \frac{s^2}{2}\sin^2\theta + ...\right)  \\
\label{eq13}
\frac{d \sigma}{d \Omega} & = & \left(\frac{d \sigma}{d \Omega}\right)_{QED}\cdot \left(1 + \frac{\pi^2}{12 (M_S)^4}\cdot \frac{s^2}{2}\sin^2\theta \right)  \\
\label{eq14}
\frac{d \sigma}{d \Omega} & = & \left(\frac{d \sigma}{d \Omega}\right)_{QED}\cdot \left(1 + \frac{(\eta_L+\eta_R)}{2 (\L^{CI})^4}\cdot \frac{s^2}{2}\sin^2\theta + ...\right)  \\
\label{eq15}
\frac{d \sigma}{d \Omega} & = & \left(\frac{d \sigma}{d \Omega}\right)_{QED}\cdot \left(1 + \frac{2}{\alpha (\L_6)^4}\cdot \frac{s^2}{2}\sin^2\theta + ...\right)  \\
\label{eq16}
\frac{d \sigma}{d \Omega} & = & \left(\frac{d \sigma}{d \Omega}\right)_{QED}\cdot \left(1 + \frac{s^3}{32\pi \alpha^2 (\L^{\prime})^6}\frac{\sin^2\theta}{1+\cos^2\theta} + ...\right).
\end{eqnarray}
The QED cut-off -~\Eqref{eq11}, is the basic form of possible deviations from
quantum electrodynamics. If we ignore higher order terms (given by $...$),
all equations except~\Eqref{eq16} predict the same form of deviations in the
differential cross section. \Eqref{eq12} is the low scale gravity
case~\cite{Giudice,Agashe},
and \Eqref{eq13} appears in the TeV strings model. The others are variations
of contact interactions. I have chosen to compile the deviations in a
similar form, which makes it particularly easy to compare the results from
different searches by transforming the relevant parameters.

\section{Fermion-pair production: $\ee \rightarrow f\bar f(\gamma)$}

\subsection*{$f\bar f$ - Signal Definition}

\hspace*{0.7cm}
The large data samples accumulated at LEP2 have propelled the
measurements of several final states in the precision area,
where many tiny theoretical details or experimental systematic
effects start to grow in importance.
For instance, the signal definition for fermion pairs is complicated by
several factors:
\begin{itemize}
\item Interplay between two-fermion and four-fermion final states:
the radiative corrections include not only effects due to vector bosons,
but also contributions from real and virtual fermion pairs,
{\it cf.}~\Figref{figure6}. For example,
low mass (secondary) fermion pairs are part of the signal, like photons.
The situation is ambiguous if the two final state fermion pairs carry the same
flavor or roughly the same energy. Fortunately these corrections are negligible
for the non-radiative (genuine high energy) sample, which is the main
search field for physics beyond the SM~\footnote{For an early review of
searches above the Z pole see {\it e.g.}~\cite{Bourilkov:1998}.}. \\
\begin{figure}[h]
\begin{center}
\resizebox{1.0\textwidth}{4.6cm}{\includegraphics*[1cm,22.4cm][9.0cm,26cm]{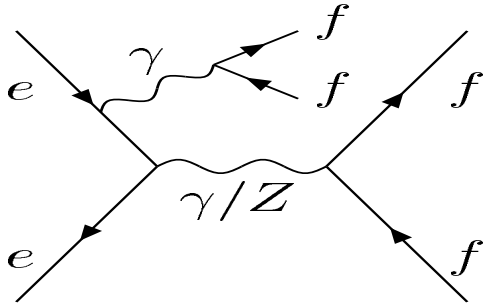}}
\end{center}
  \caption{\em One example of Feynman graph contributing to pair-corrections
           in fermion-pair production.}
  \label{fig:figure6}
\end{figure}
\item  The definition of the effective (propagator in~\Figref{figure1}) energy
       $\SP$ is not unique due to interference between initial- and final-state
       radiation:
 \begin{itemize}
       \item  the interference changes the cross-sections: 
       for $\sqrt{\SP} / \sqrt{s} > 0.85$ by $\sim$ 1.5\% for $\MM$
       and $\sim$ 0.5\% for $\qq$ in order $\cal O (\alpha)$
      (the experimental efficiency is changed by \mbox{$\leq$ 50\%} of this)
       \item  the experiments either apply a correction by subtracting the
       interference contribution, rendering the $\sqrt{\SP}$ definition unique,
       {\em or}
       define $\sqrt{\SP}$ as the effective mass of the outgoing fermion pair
       after final state radiation; there is no unanimity about the ``best''
       definition, requiring the minimal amount of theoretical corrections;
       for a detailed discussion we refer the reader to~\cite{LEP2MC}.
 \end{itemize}
\item \mbox{ Rising four-fermion background}: LEP2 is the kingdom of double
       (or single) weak boson production like W$^+$W$^-$, ZZ, Z$\EE$,
       We$\nu$, ...
\end{itemize}

\subsection*{Fermion Pairs - Selection}

\begin{figure}[!p]
\begin{center}
\resizebox{0.49\textwidth}{9.00cm}{\includegraphics{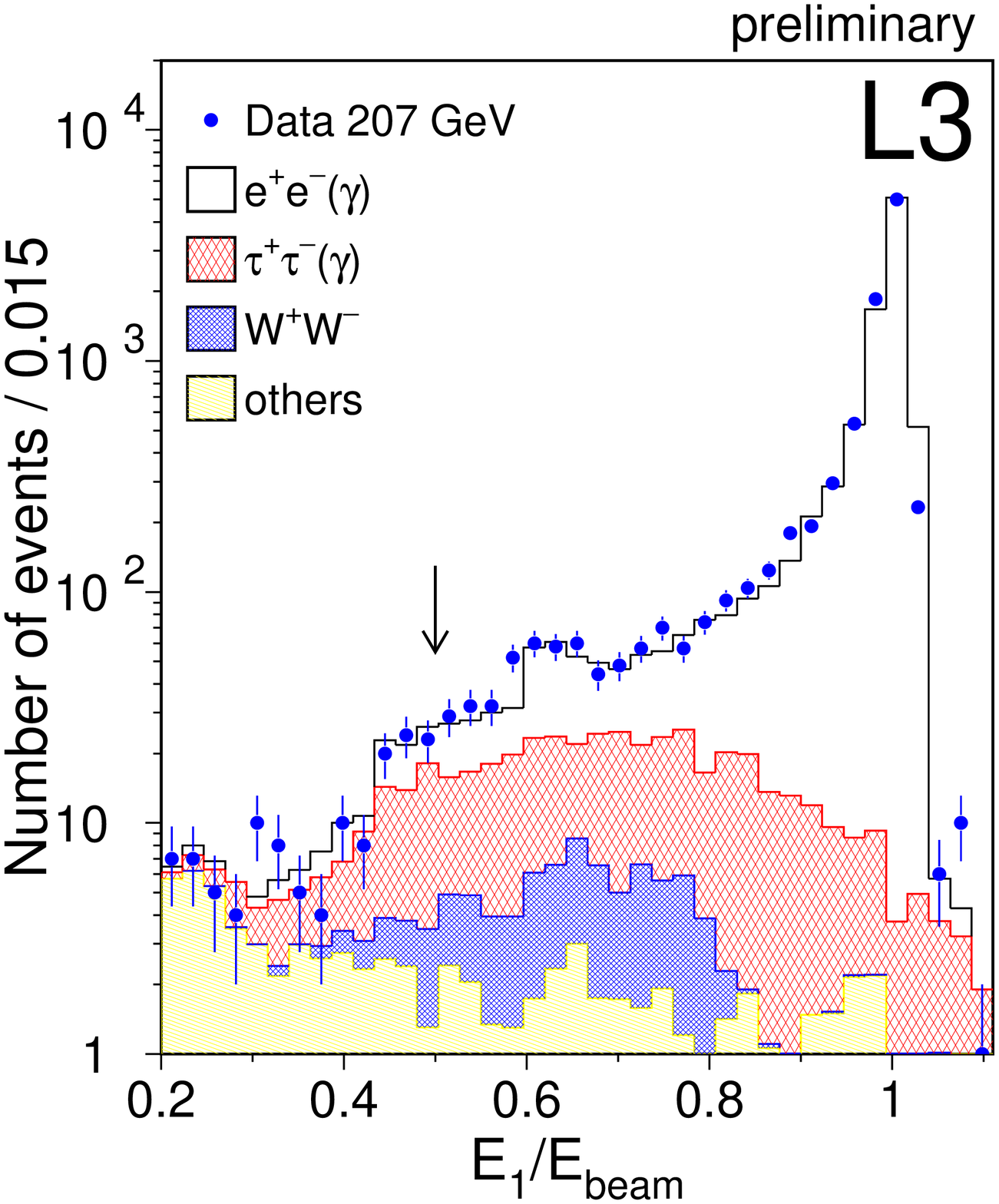}}
\resizebox{0.49\textwidth}{9.00cm}{\includegraphics{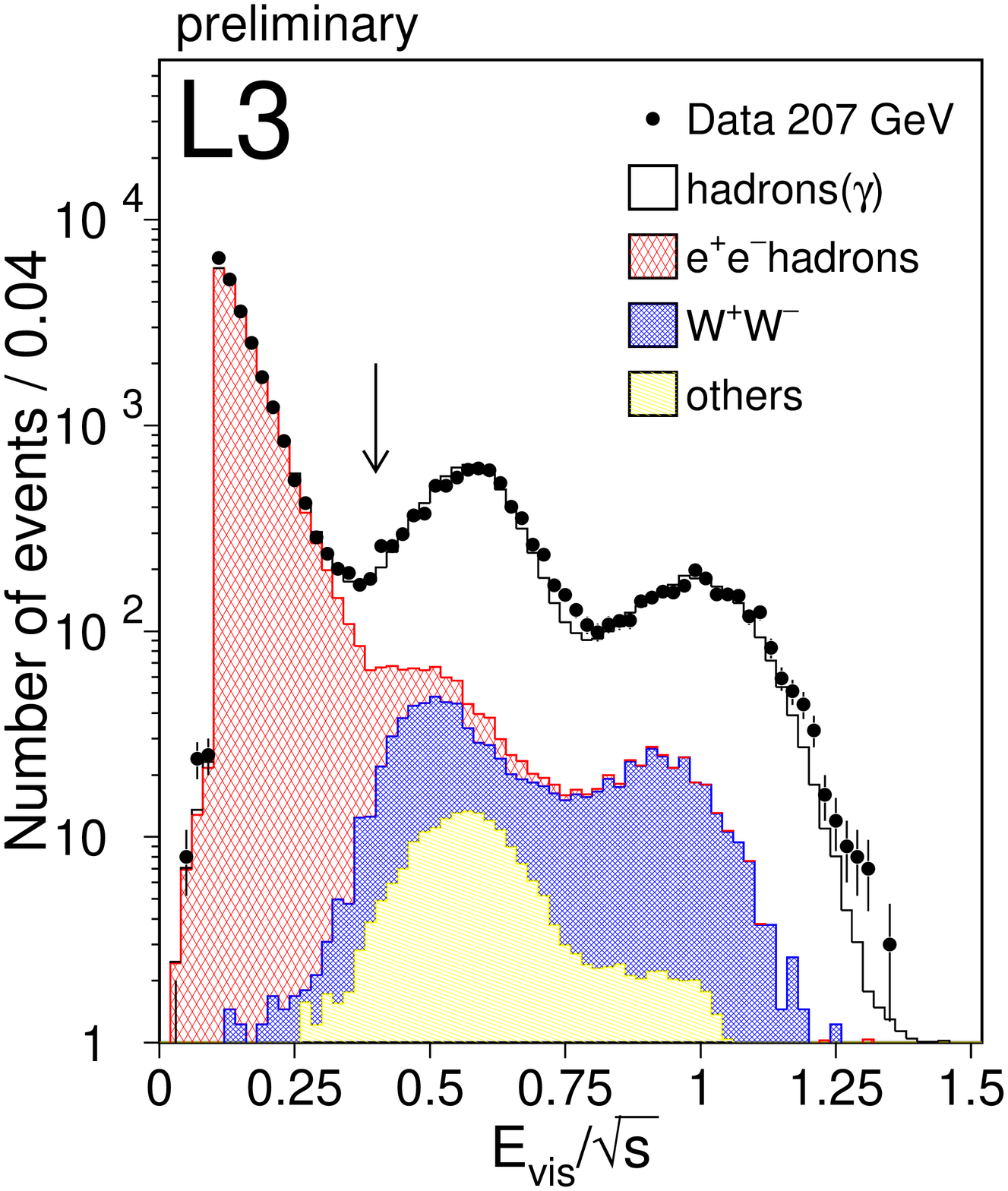}}
\resizebox{0.49\textwidth}{9.00cm}{\includegraphics{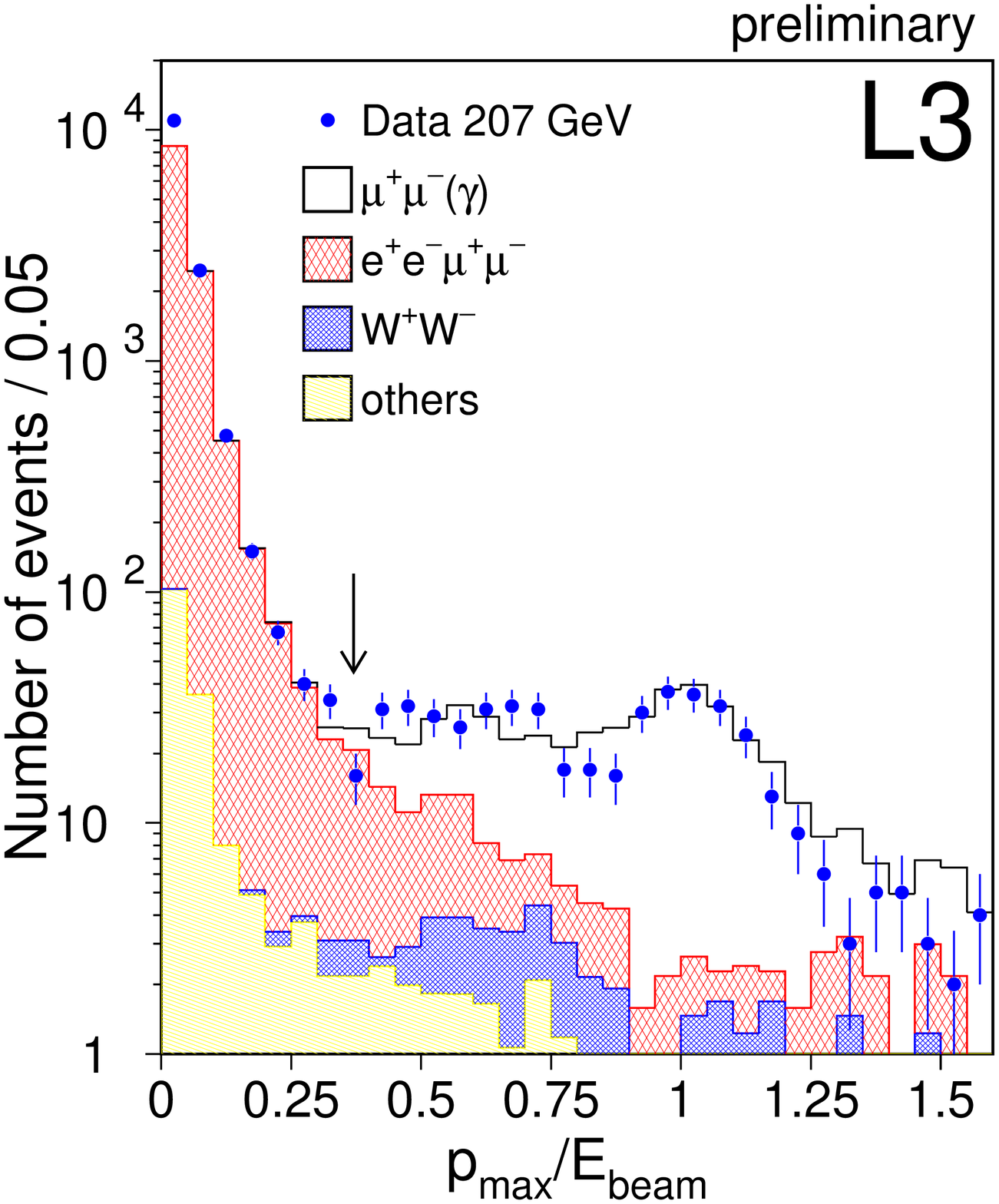}}
\resizebox{0.49\textwidth}{9.00cm}{\includegraphics{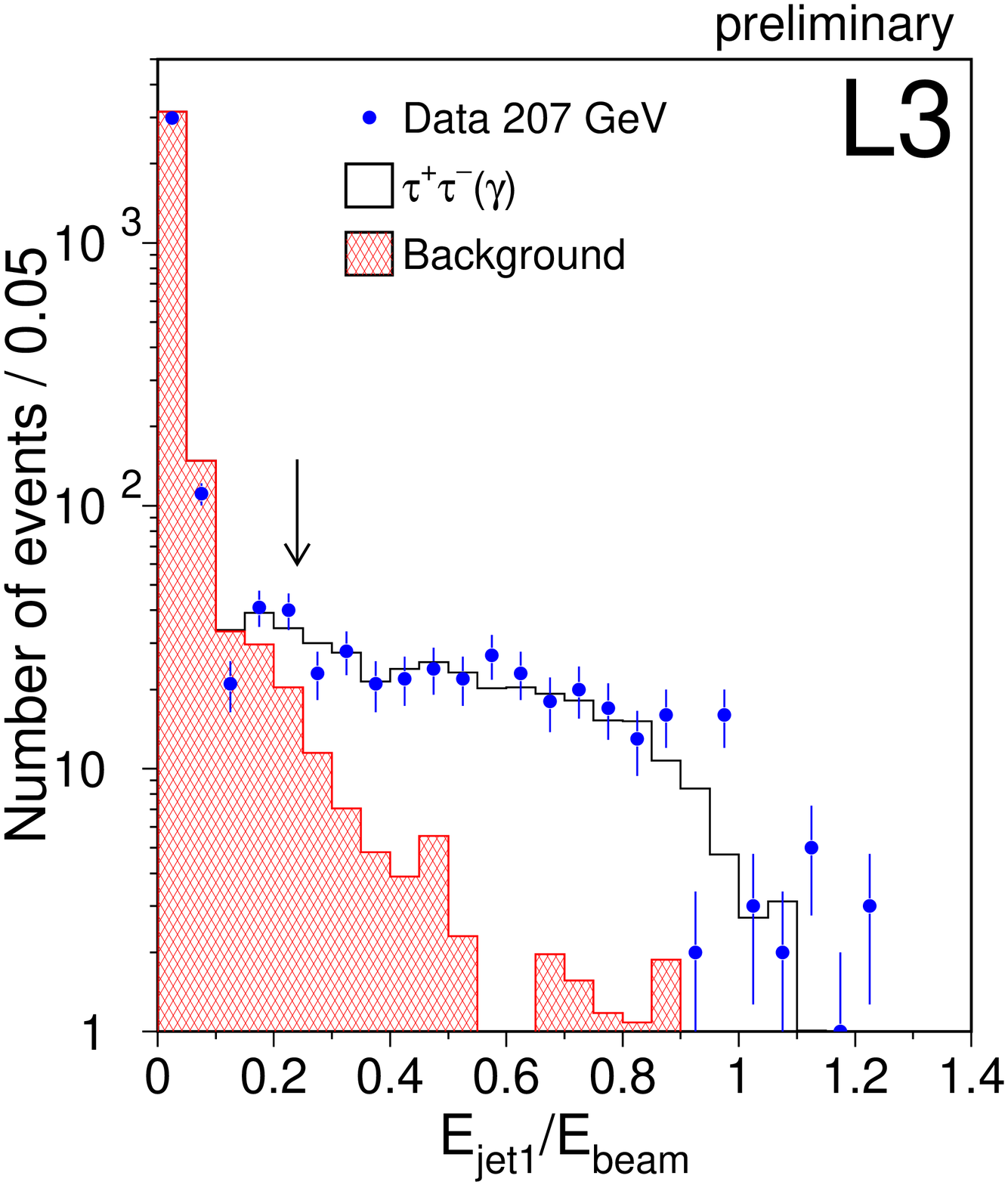}}
\end{center}
  \caption{\em Typical plots for the event selection of the various
   final states ($\ee$, $\qq$, $\MM$, $\TT$) in the L3 experiment.}
  \label{fig:figure7}
\end{figure}

\hspace*{0.7cm}
The event selection for the different final states is illustrated
on~\Figref{figure7}.
For $\ee$ it is `quite easy', and consists of requiring an
identified electron-positron pair (typically tracks matched to
electromagnetic energy deposits).
The huge peak at electron (or positron) energy in the vicinity
of the beam energy is due to the large contribution form $t$-channel
photon exchange. The backgrounds are tiny.
For $\qq$ the selection is `easy', and asks for high multiplicity
relatively balanced events with substantial visible energy.
The high energy peak is accompanied by a huge peak due to radiative
`Z~returns', favored by the large hadron cross section at the Z pole.
For $\MM$ the selection is `still easy', asking for two muons identified
as tracks in the muon chambers or as minimum ionizing objects in the
calorimeters. The ratio of the muon and beam
momenta clearly separates the signal from various backgrounds peaking at
low energies. Cosmic muons complicate the picture, and are typically
suppressed by imposing space-time constrains on the tracks to originate
from beam interactions.

\hspace*{0.7cm}
For $\TT$ the selection is `far from easy', and consists typically of
requiring two narrow low multiplicity jets. The ratio of jet to beam
energy suppresses low energy backgrounds, and the peak at one is
absent due to the neutrinos from $\tau$ decays. Bhabha backgrounds
are dangerous, especially if there are problems with energy measurements
in the endcaps.

\subsection*{LEP2 $f\bar f$ Group}

\hspace*{0.7cm}
Following the LEP tradition, a Fermion-pair Subgroup of the LEP Electroweak
Working Group has been established in 1999 in order to 
combine the measurements of total and differential cross sections
$\sigma_{tot}$ and $\frac{d \sigma}{d \COST}$, of forward-backward 
asymmetries $\AFB$, and of the ratios of heavy flavor to total
hadronic cross sections for beauty and charmed quarks $R_b$ and $R_c$.
In this way the statistical error of the combined measurements is reduced
by a factor of two, and several types of correlated systematic errors, {\it e.g.}
in one experiment between different centre-of-mass energies or between all
experiments, can be properly taken into account. The group also provides
a very useful forum for exchange of ideas and facilitates the convergence
of the signal definitions used by the different collaborations, making
the task of combining the measurements much easier.

At the moment of this writing we have succeeded to combine the 
published results~\mcite{al183,de172,*de189,l3136,*l3172,*l3189,*l3rb189,
*sneut,*l3ci,*l3gr1,*l3gr2,*l3189newph,op172,*op183,*op189}
and preliminary measurements~\mcite{al189,*al202,*al209,de202,*de209,
l3202,*l3209,op202,*op209}
for the following final states:
$\MM$, $\TT$, $\qq$, $\rm b\bar b$ and $\rm c\bar c$.
The results~\cite{LEP2ff} are shown in the next subsection.
They are used to perform combined fits searching for contact interactions 
or for Z$^{\prime}$.

The future plans of the group include combining the measurements of the
$\EE$  and  $\GG$  final states. In the absence of LEP--wide combined
measurements, for the purpose of this talk I have followed a top-down
approach and extracted limits on models beyond the Standard Model
by fitting directly the preliminary results of the four LEP collaborations
for $\ee$, or by using as input the results of the fits for the 
QED cut-off of the individual experiments for $\GG$.

\subsection*{Results}

\begin{figure}[!p]
\begin{center}
\resizebox{0.49\textwidth}{11.0cm}{\includegraphics{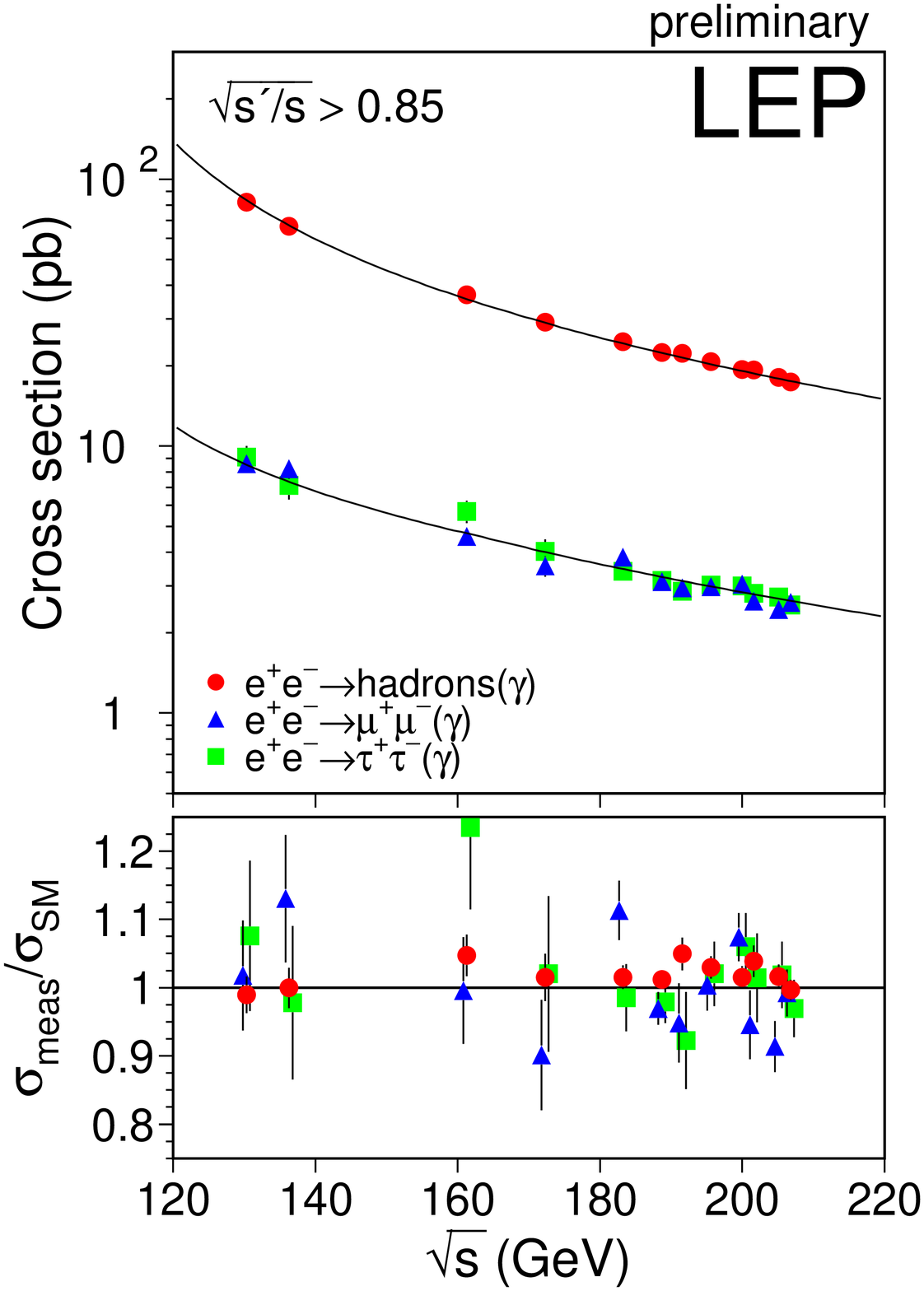}}
\resizebox{0.49\textwidth}{11.0cm}{\includegraphics{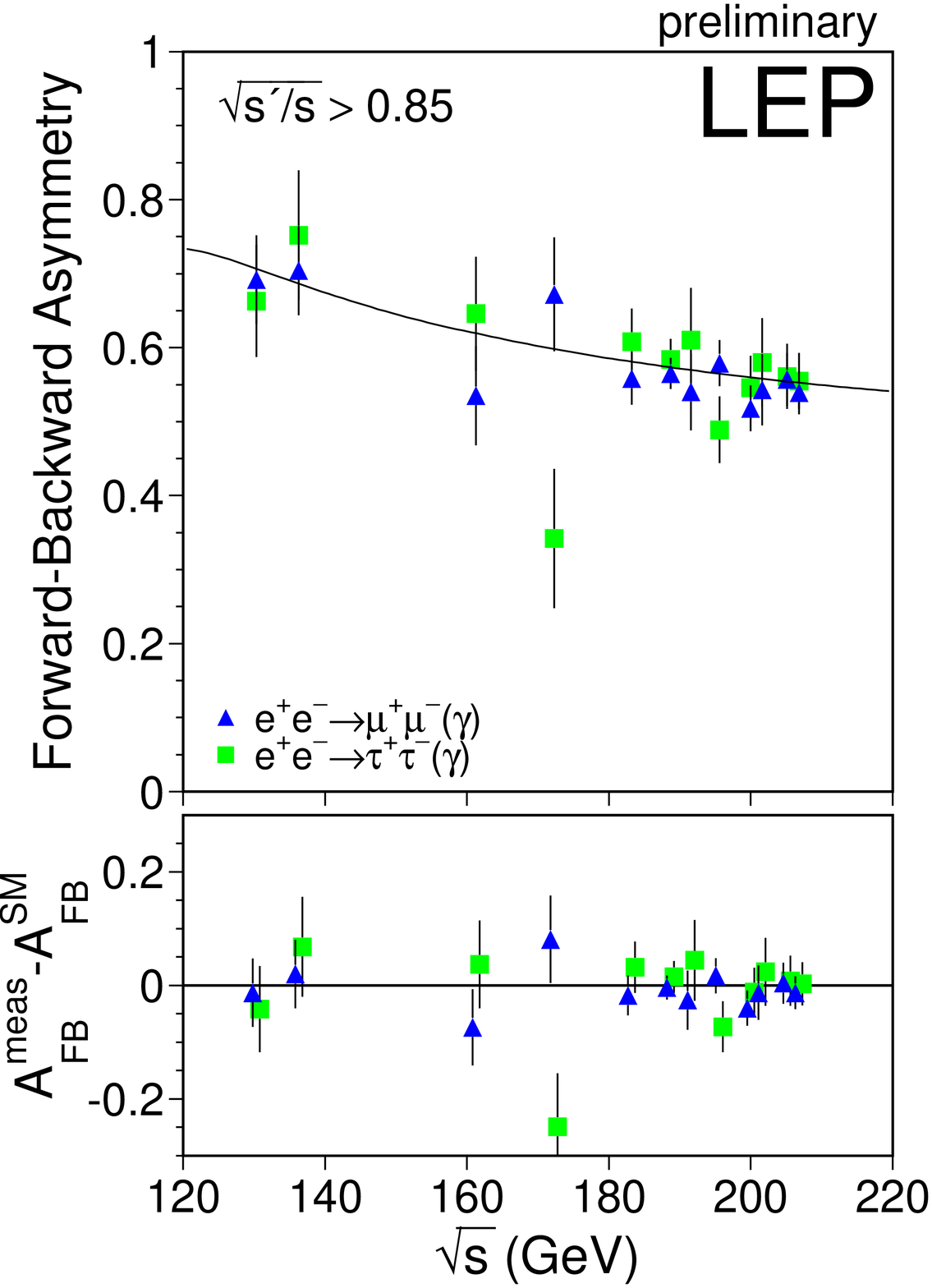}}
\resizebox{0.49\textwidth}{11.0cm}{\includegraphics{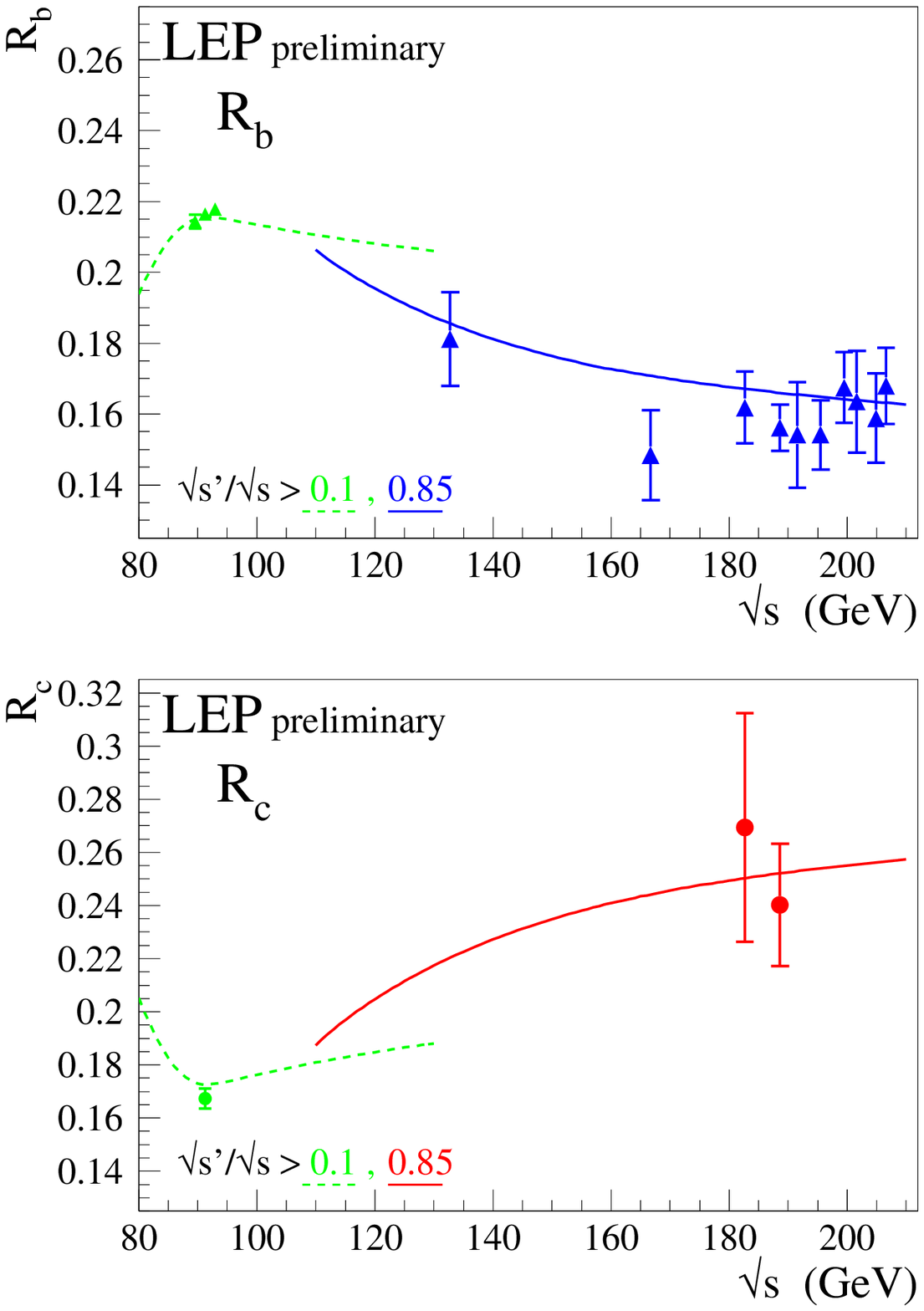}}
\resizebox{0.49\textwidth}{11.0cm}{\includegraphics{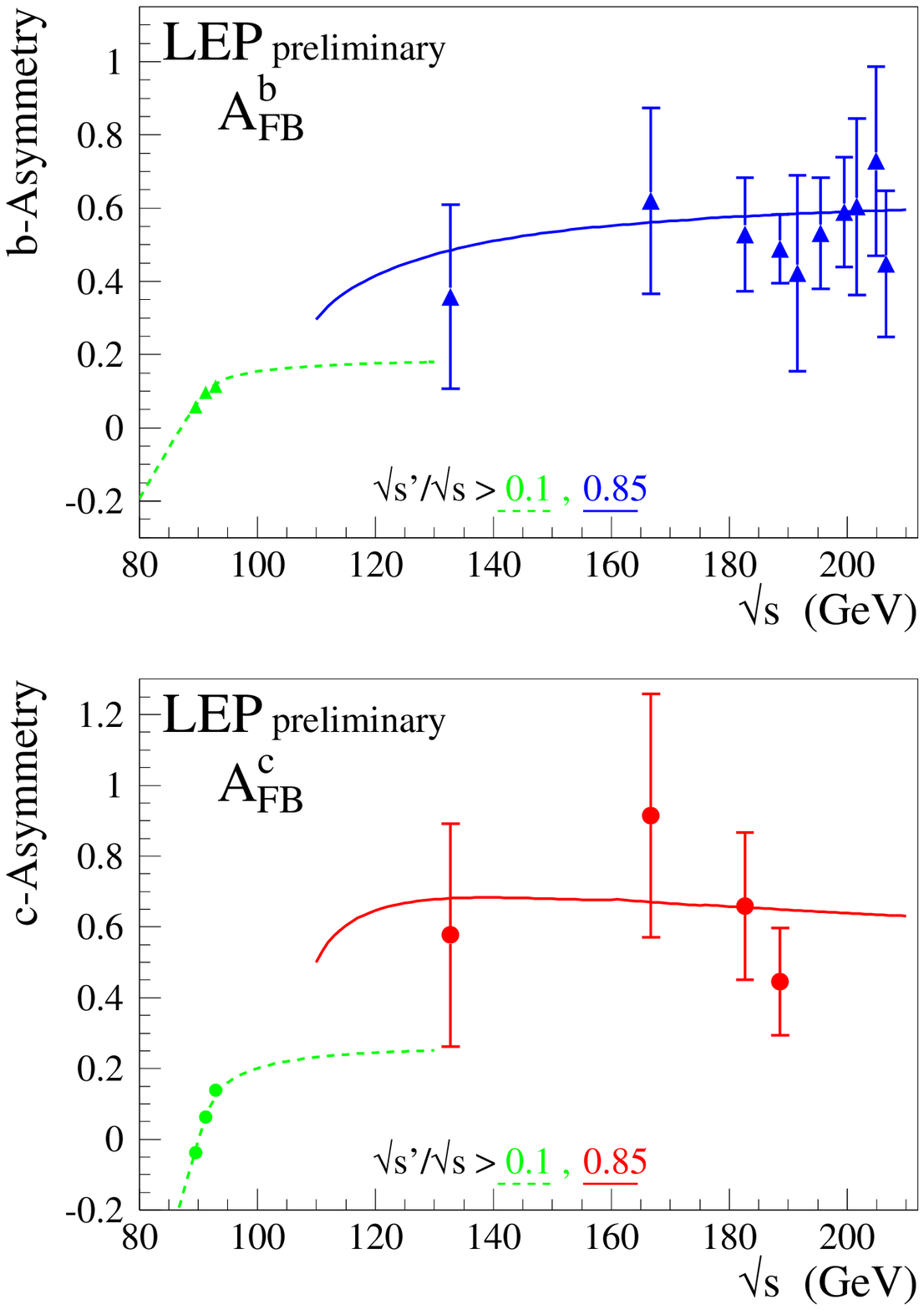}}
\end{center}
  \caption{\em Total cross sections and forward-backward asymmetries for
           $\MM$, $\TT$ and $\qq$ (upper half). $R_b$ and $\AFB$
           for $\rm b$ (middle), $R_c$ and $\AFB$ for $\rm c$ (lower quarter) quarks.}
  \label{fig:figure8}
\end{figure}

\hspace*{0.7cm}
The combined results for the total cross section $\sigma_{tot}$ and
for the forward-backward asymmetry $\AFB$ are shown in~\Figref{figure8}.
In the same figure the measurements for $\rm b$ and $\rm c$ quarks are shown.
They agree well with the SM expectations above the Z resonance.
A special case is the hadron cross section, which shows the tendency
to lie above the SM line, computed with ZFITTER~\cite{ZFITTERNPH}.
The averaged cross section is between 2 and 3 standard deviations
away from the expectation, depending on the assigned normalization
uncertainty. The high precision of the measurements requires improvements
in the theoretical uncertainty and independent cross-checks.

%\vspace{-1.2cm}

%\newpage
\section{Photon-pair production: $\ee \rightarrow \GGG$}

\begin{figure}[!p]
\begin{center}
\resizebox{0.48\textwidth}{11.0cm}{\includegraphics*[0cm,6cm][34.0cm,28.0cm]{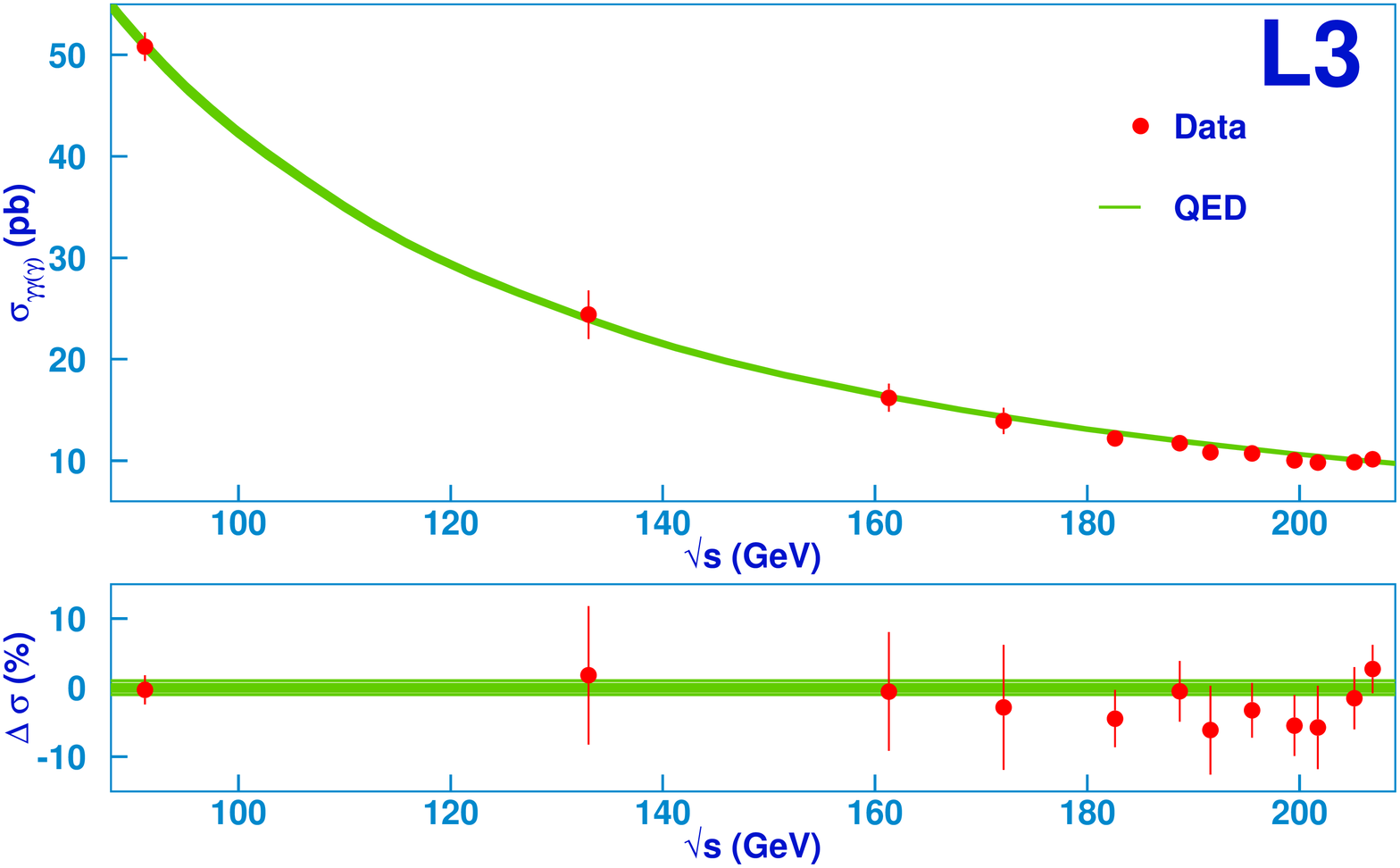}}
\resizebox{0.48\textwidth}{11.0cm}{\includegraphics*[0cm,0cm][15.0cm,18.0cm]{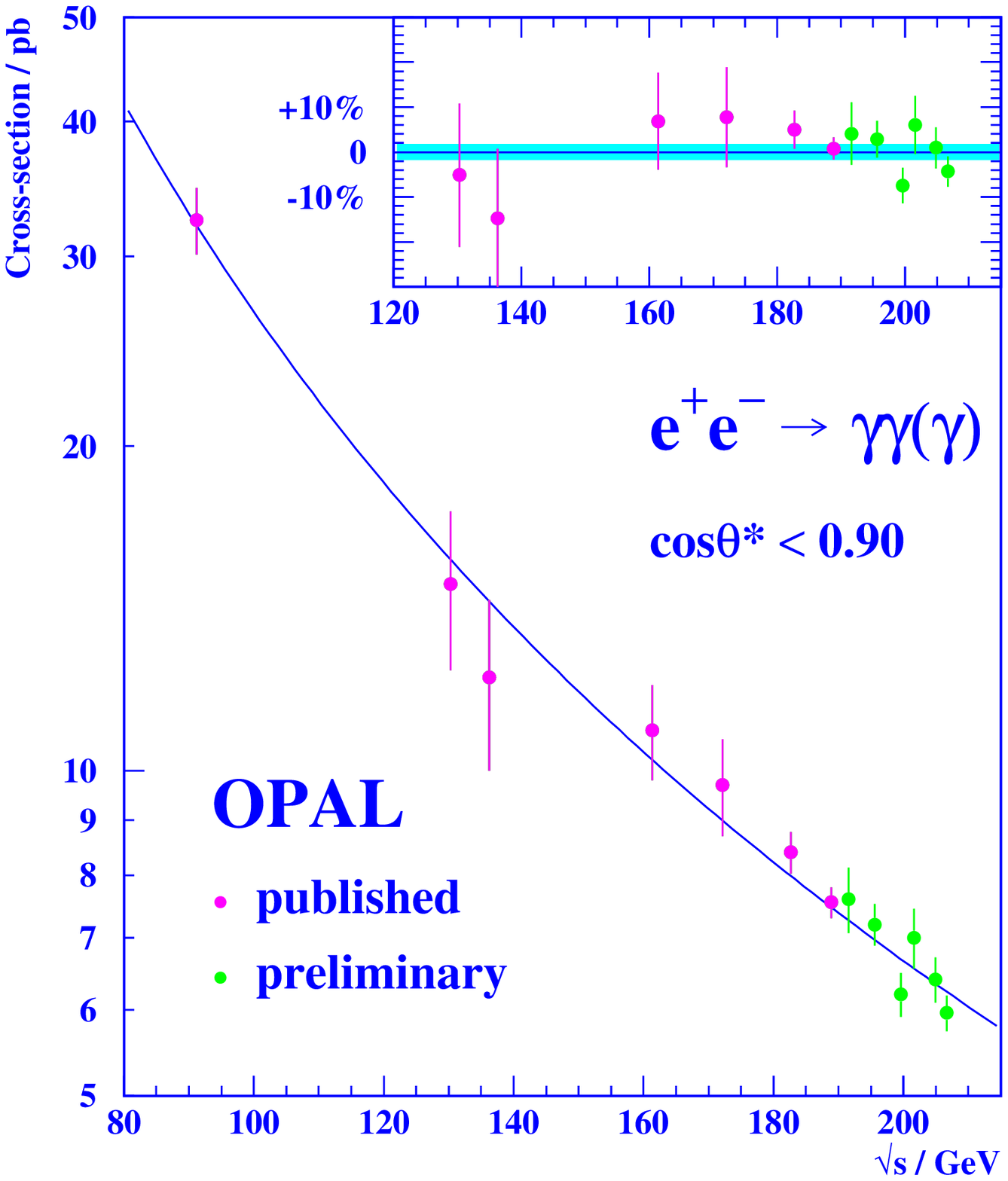}}
  \vspace{-0.6cm}
\resizebox{0.48\textwidth}{10.0cm}{\includegraphics{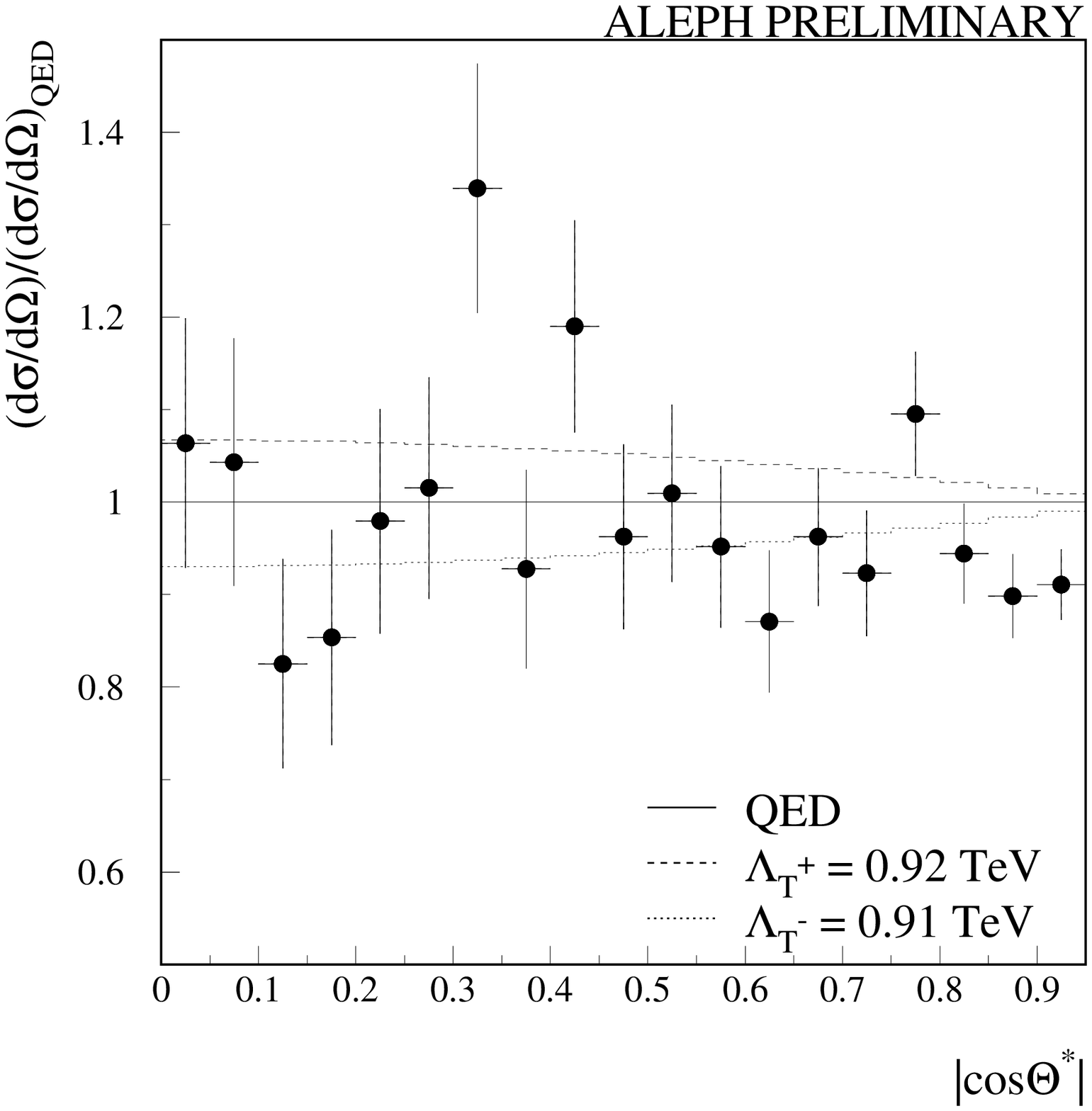}}
\resizebox{0.48\textwidth}{10.0cm}{\includegraphics{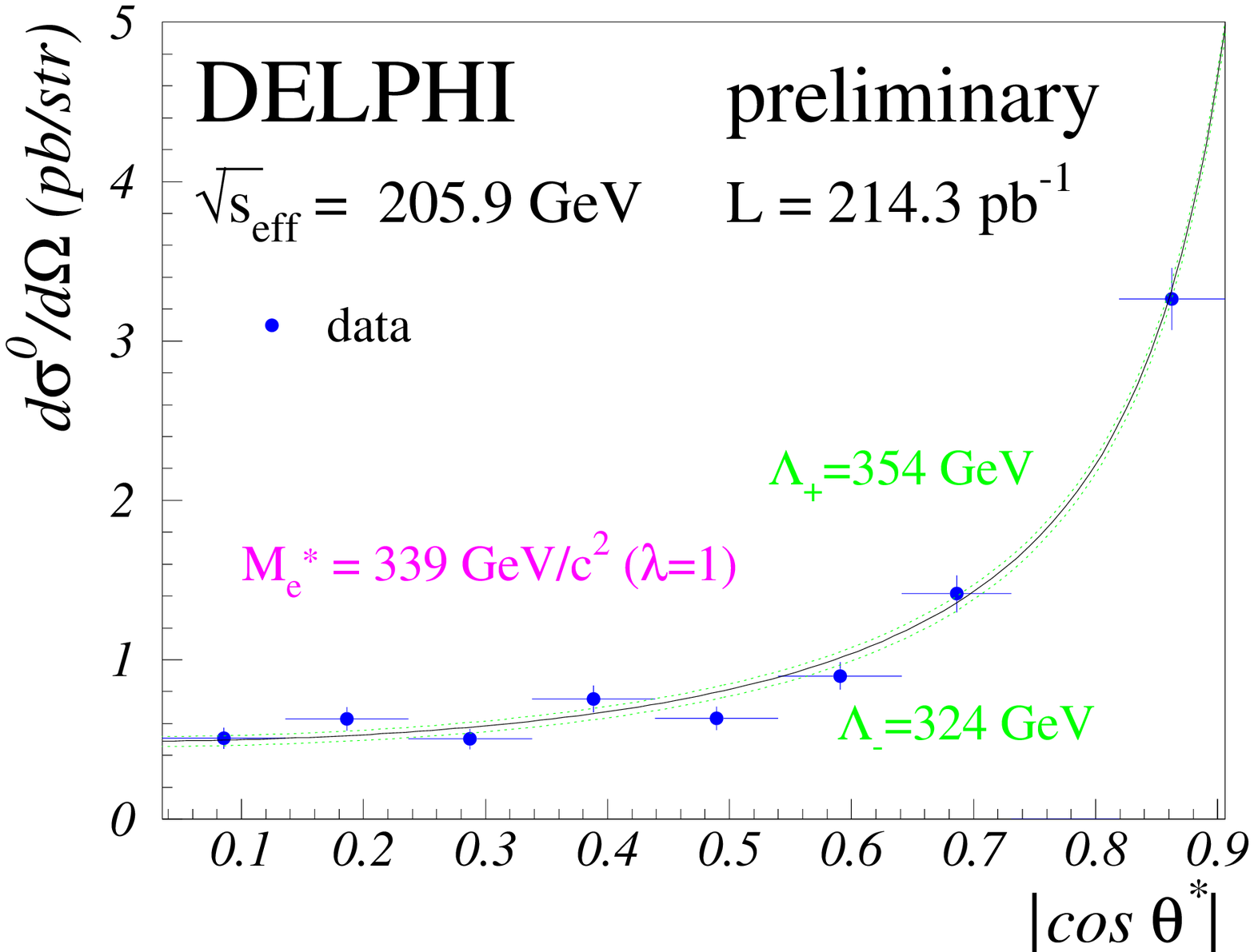}}
\end{center}
  \caption{\em Total cross section measurements for diphoton final states from
           L3 and OPAL.
           Ratio of the differential cross section to the SM from ALEPH and
           the effects expected in low scale gravity models.
           Differential cross section measurements from DELPHI and the 
           fit results for QED cut-offs and excited electrons.}
  \label{fig:figure9}
\end{figure}

\hspace*{0.7cm}
The event selection is similar to the one for $\ee$, with the major
difference that here the (absence of) tracking information is used
to veto tracks associated with the electromagnetic clusters,
and there is no need to recognize the electric charge.
Measurements of total and differential cross sections for {$\GG$} final
states~\cite{al202gg,*de209gg,*l3209gg,*op209gg} are shown in~\Figref{figure9}.
The results include the 2000 data up to 209 GeV (for ALEPH they
are up to 202 GeV). Good agreement with the SM expectations is observed.

%\vspace{-1.2cm}

\newpage
\section{Searches for New Phenomena: Contact Interactions and Extra Dimensions}

\begin{figure}[!ht]
  \begin{center}
  \begin{tabular}{cc}
     \resizebox{0.48\textwidth}{10.9cm}{\includegraphics{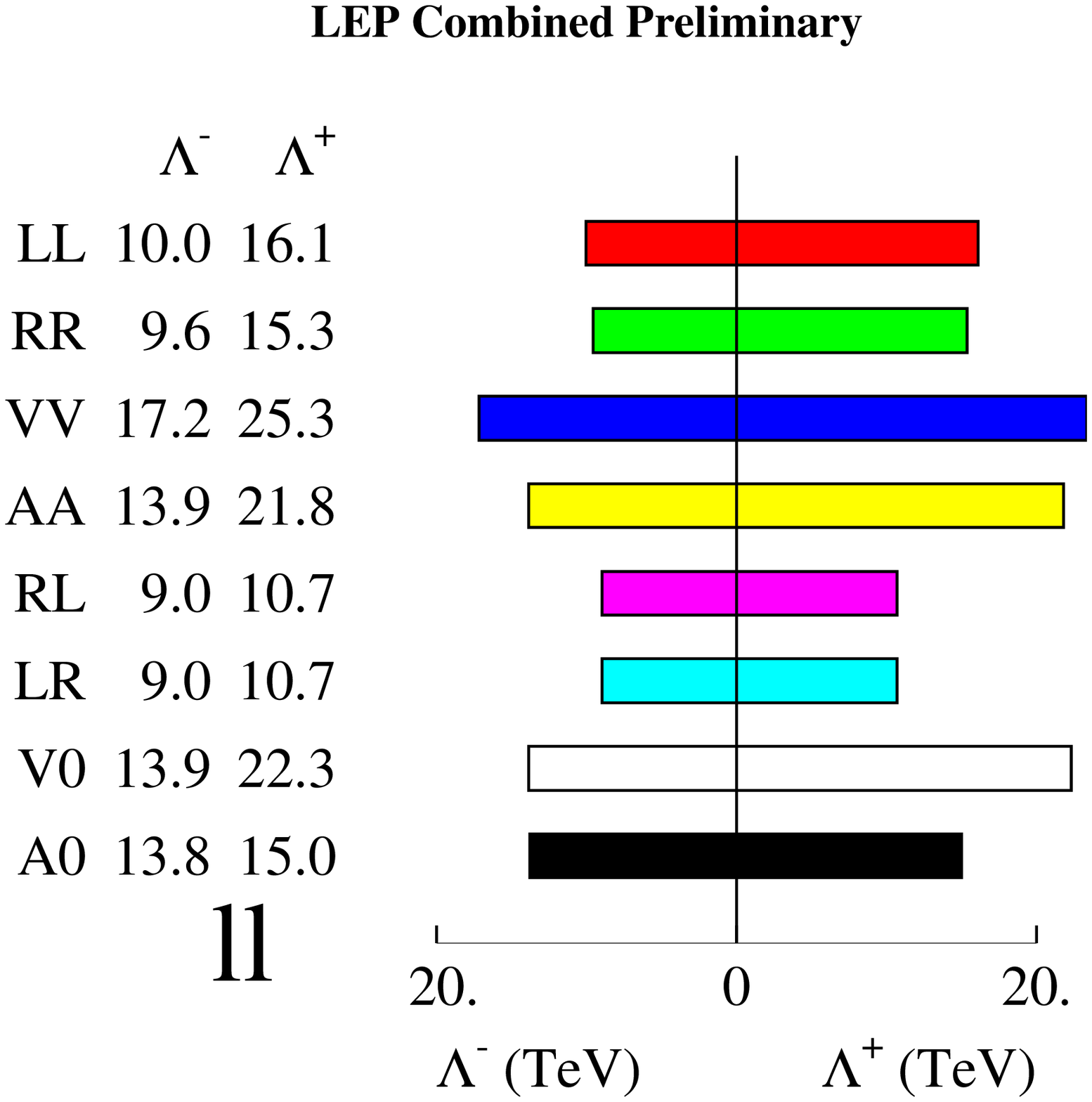}} &
     \resizebox{0.48\textwidth}{10.9cm}{\includegraphics{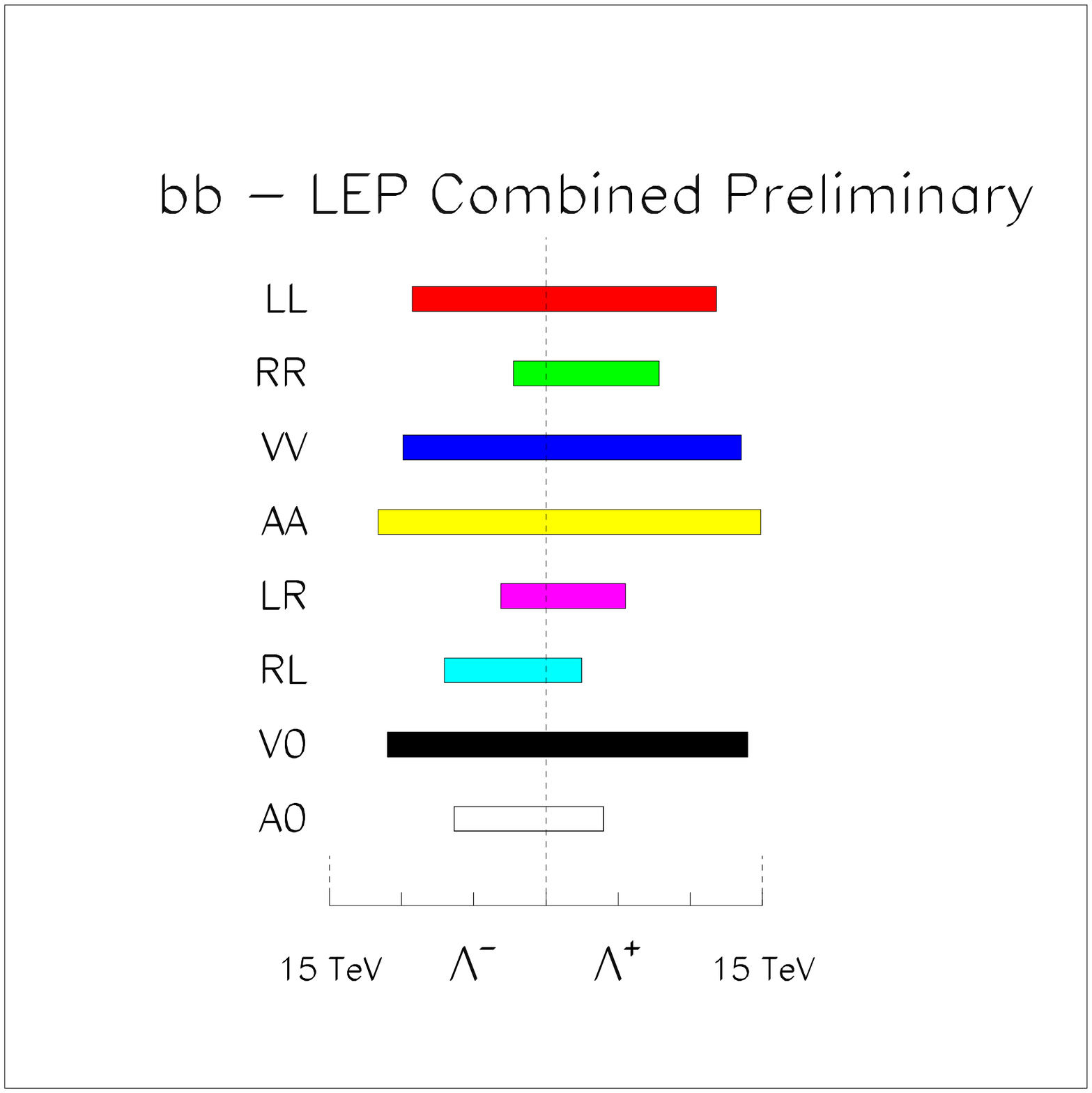}} \\
  \end{tabular}
  \end{center}
  \caption{\em Limits on contact interactions at 95\% CL.
            a - for $\MM$ and $\TT$ combined ($ll$), b - for $b\bar b$.}
  \label{fig:figure10}
\end{figure}

\vspace{-0.3cm}
\hspace*{0.7cm}
The LEP measurements show no statistically significant deviations from the SM
predictions. In their absence we present limits at 95 \% confidence level
on several new physics proposals.
The LEP2 $f\bar f$ group has derived limits on contact
interactions~\cite{LEP2ff}
for  $\MM$, $\TT$ and the two channels combined (\Tabref{ci-limits}, 
\Figref{figure10}a),
for $\qq$, $b\bar b$ (\Figref{figure10}b) and $c\bar c$.

%%%%%%%%%%%%%%%%%%%%
%% Values for 0.5% theory error
\vspace{-0.4cm}
\begin{table}[ht]
\renewcommand{\arraystretch}{1.05}
\caption{Limits on contact interactions at 95\% CL.}
\label{tab:ci-limits}
 \begin{center}
% mm tt
  \begin{tabular}{|c|r|c|c||r|c|c|}
   \hline
   \multicolumn{4}{|c||}{\boldmath $e^{+}e^{-} \rightarrow \mu^{+}\mu^{-}$\unboldmath} & \multicolumn{3}{c|}{\boldmath $e^{+}e^{-} \rightarrow \tau^{+}\tau^{-}$\unboldmath} \\
   \hline
   Model  & $\varepsilon$    &  $\Lambda^{-}$  & $\Lambda^{+}$ & $\varepsilon$    &  $\Lambda^{-}$ & $\Lambda^{+}$ \\
          & $[\TeV^{-2}]$ &  $[\TeV]$       & $[\TeV]$      & $[\TeV^{-2}]$ &  $[\TeV)]$     & $[\TeV]$      \\
   \hline
   \hline
   LL & -0.0056$^{+ 0.0040}_{- 0.0040}$ &    8.6 &   14.7 & -0.0016$^{+ 0.0054}_{- 0.0047}$ &    9.4 &   10.6 \\
   \hline                                                                                                       
   RR & -0.0077$^{+ 0.0053}_{- 0.0030}$ &    8.3 &   14.1 & -0.0008$^{+ 0.0049}_{- 0.0060}$ &    9.0 &   10.2 \\
   \hline                                                                                                       
   VV & -0.0014$^{+ 0.0016}_{- 0.0017}$ &   15.3 &   22.4 & -0.0002$^{+ 0.0016}_{- 0.0023}$ &   15.2 &   17.5 \\
   \hline                                                                                                       
   AA & -0.0036$^{+ 0.0027}_{- 0.0013}$ &   11.8 &   20.4 & -0.0004$^{+ 0.0032}_{- 0.0025}$ &   13.2 &   13.8 \\
   \hline                                                                                                       
   LR &  0.0014$^{+ 0.0043}_{- 0.0061}$ &    7.8 &    9.3 & -0.0014$^{+ 0.0075}_{- 0.2283}$ &    2.1 &    8.6 \\
   \hline                                                                                                       
   RL &  0.0014$^{+ 0.0043}_{- 0.0061}$ &    7.8 &    9.3 & -0.0014$^{+ 0.0075}_{- 0.2283}$ &    2.1 &    8.6 \\
   \hline                                                                                                       
   V0 & -0.0036$^{+ 0.0025}_{- 0.0014}$ &   12.1 &   20.2 & -0.0003$^{+ 0.0023}_{- 0.0030}$ &   13.0 &   14.7 \\
   \hline                                                                                                       
   A0 &  0.0008$^{+ 0.0020}_{- 0.0031}$ &   12.7 &   12.9 & -0.0008$^{+ 0.0038}_{- 0.0046}$ &    9.9 &   11.9 \\
   \hline
  \end{tabular}
 \end{center}
\end{table}
\vspace{-0.4cm}

For many models the limits are in the 10--20 TeV range.
It is interesting to compare them with the expectations back in
1988~\cite{Schrempp:1988}. The actual limits are typically a factor two
better, mainly due to the successful running of LEP above the
anticipated maximum energy of 190 GeV. Another factor is the combination of the
results from the four collaborations.

The values for $\varepsilon$ from~\Tabref{ci-limits} are maximally 1.5 standard
deviations away from the Standard Model value $\varepsilon = 0$.
The recent measurement from the E821 experiment at BNL~\cite{E821} on the
anomalous magnetic moment of the muon shows a 2.6 standard deviations
discrepancy with theory. As pointed out in~\cite{Lane}, if the experimental
result is ascribed to muon substructure, the model-independent limit on the
energy scale is $1.2\; \TeV < \L_{\mu} < 3.2\; \TeV$ at 95 \% CL.
The LEP limits from $\ee \rightarrow \MM$ are based on the assumption
of a common energy scale for electrons and muons: $\L_e \simeq \L_{\mu}$.
While there is no {\it a priori} reason for the first- and second-generation
lepton substructure scales to be close, the 10--20 TeV range of the
LEP limits implies a very high scale for electrons if the new experimental
result is originating from muon compositeness.

\begin{figure}[!b]
\begin{center}
  \begin{tabular}{cc}
    \resizebox{0.48\textwidth}{11.0cm}{\includegraphics*[0cm,0cm][20.0cm,18.0cm]{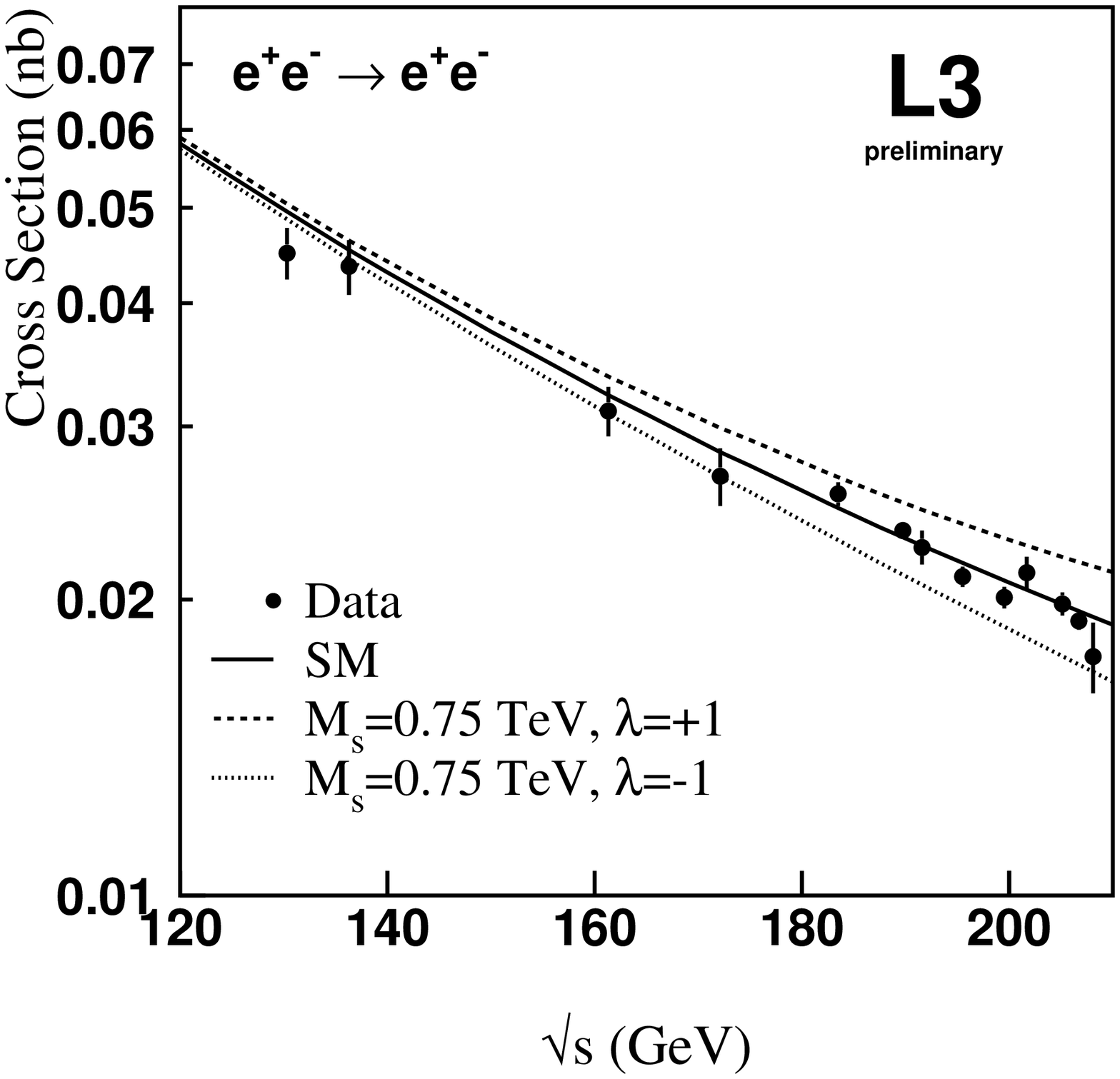}}
    \resizebox{0.48\textwidth}{11.0cm}{\includegraphics*[0cm,0cm][18.0cm,23.0cm]{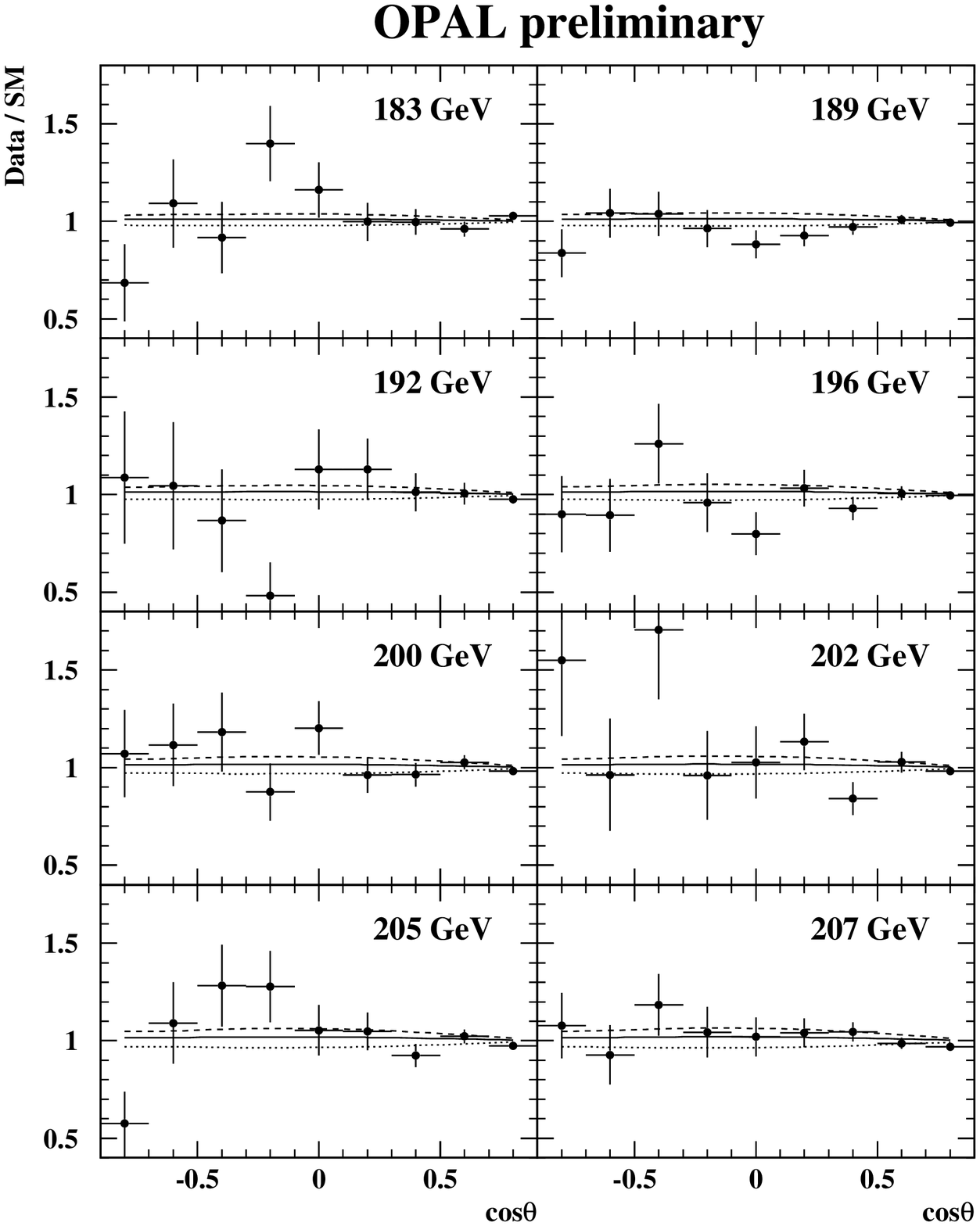}}
%   \resizebox{0.48\textwidth}{11.0cm}{\includegraphics{lsg-l3.eps}}
%   \resizebox{0.48\textwidth}{11.0cm}{\includegraphics{lsg-opal.eps}}
  \end{tabular}
\end{center}
  \caption{\em Measurements of the reaction $\ee \rightarrow \ee$ and searches
           for low scale gravity effects: left - total cross sections from L3,
           right - differential cross sections from OPAL.}
  \label{fig:figure11}
\end{figure}

Some experiments~\cite{l3209newph,*op209lsg} have updated their limits on the
low gravity or TeV string scales, using the most sensitive channel at LEP2 -
Bhabha scattering~\cite{Bourilkov:1999,Bourilkov:2000}.
The measurements are compared to the SM or to theories with large extra
dimensions in~\Figref{figure11}. The results are:
\Be
{\rm L3}: M_s^+ > 1.06\;\TeV \ \ M_s^- > 0.98\;\TeV\ \ \ \ TeV\ Strings: M_S > 0.57\;\TeV
\Ee
\Be
{\rm OPAL}: M_s^+ > 1.00\;\TeV \ \ M_s^- > 1.15\;\TeV.
\Ee
%\vspace{-1.0cm}

\newpage
The reaction $\ee \rightarrow \GG$ is a sensitive ``QED laboratory''.
The limits obtained by the LEP
experiments~\cite{al202gg,*de209gg,*l3209gg,*op209gg} are compiled
in~\Tabref{gg-limits}. In the last column of the table I have
combined the limits, using as input the results of the fits for the 
QED cut-off of the individual collaborations. Here the assumption
that the log-likelihood curve in the variable $1/\L^4$ can be
approximated with one coming from the normal distribution is made.
Then it is straightforward to combine the measurements and to
extract LEP-wide limits.
The individual results are very close to this behaviour, and the
resulting uncertainty in the limits is only a few~\%.
This is used further as input to translate the limits in the language
of low scale gravity and TeV strings, taking into account also the sign
conventions.

\begin{table}[!ht]
\renewcommand{\arraystretch}{1.20}
\caption{Limits on new physics from $\GG$ at 95 \% CL.}
\label{tab:gg-limits}
  \begin{center}
    \begin{tabular}{|l||c|c|c|c||c|}
       \hline
 Mo-           &  ALEPH   &  DELPHI  &   L3     &   OPAL   &  Combined  (DB)\\
 del           & $[\TeV]$ & $[\TeV]$ &$\ [\TeV]\ $ & $[\TeV]$ & $[\TeV]$ \\
       \hline                  
\multicolumn{6}{|c|}{QED cut--off - $\L_{\pm}^{QED}$} \\
       \hline                  
$\L_+$         &  0.319   &  0.354   &  0.385   &  0.344   &    0.44  \\
$\L_-$         &  0.317   &  0.324   &  0.325   &  0.325   &    0.37  \\
       \hline
\multicolumn{6}{|c|}{excited electron: $m_{e^*}^2/\lambda \sim (\L^{QED})^2$} \\
       \hline                  
$M_{e^*}$      &  0.337   &  0.339   &  0.325   &  0.354   &    -      \\
       \hline                  
\multicolumn{6}{|c|}{low scale gravity: $M_s^{\pm} = 2.57\;\L^{QED}_{\mp}; \L_T = 1.1195\;M_s$} \\
       \hline                  
$M_s^+$        &  0.81    &  0.83    &  0.83    &  0.83    &    0.95  \\
$M_s^-$        &  0.82    &  0.91    &  0.99    &  0.89    &    1.14  \\
       \hline
\multicolumn{6}{|c|}{TeV strings: $M_S = 0.952\;\L^{QED}$} \\
       \hline                  
$M_S$          &  0.304   &  0.337   &  0.367   &  0.327   &    0.42  \\
       \hline                  
\multicolumn{6}{|c|}{contact interactions: $\L_6 = 4.069\;\L^{QED}$} \\
       \hline                  
$\L_6$         &  1.299   &          &  1.566   &          &           \\
       \hline                  
\multicolumn{6}{|c|}{more contact interactions: $\L^{\prime}$} \\
       \hline                  
$\L^{\prime}$  &          &          &  0.810   &  0.763   &           \\
       \hline
    \end{tabular}
  \end{center}
% $^*$ Combined by DB  \\
% {\Large $\L_T = 1.1195\;M_s$} 
\end{table}

The need to obtain combined limits on the low gravity scale has been recognized
early~\cite{Gupta,*Cheung:1999,Bourilkov:1999,SMele:2000},
and LEP results are shown in~\Tabref{lsg-limits},
together with the new L3 and OPAL limits
presented for this conference. In the last part of the table the results of
combined fits to the preliminary measurements of $\sigma_{tot}$ and $\AFB$
for Bhabha scattering are shown. Here I have updated the
analysis~\cite{Bourilkov:1999}, applying the same technique. The highest
energy LEP data on $\ee$ and $\GG$ clearly improve the sensitivity
of the first combined analyses, which used data up to 189 GeV.
Applying the method for combining the $\GG$ limits, described above,
a LEP--wide limit, shown in the last row of~\Tabref{lsg-limits},
is extracted from the two most sensitive channels.

It is interesting to compare the newest LEP limits with the results from
other colliders. The D$\O$ collaboration at FNAL has performed an
analysis~\cite{d0lsg} using events containing a pair of electrons
or photons, which turn out to give the highest sensitivity at the
TEVATRON. The two-dimensional distribution in invariant mass and
polar angle is fitted. The result is:
\Be
{\rm D}\O: M_s^+ > 1.1\;\TeV \ \ M_s^- > 1.0\;\TeV.
\Ee
It is interesting to note that in spite of the low statistics at
high mass, the extended energy coverage of the hadron collider
is helpful, and the limits are pretty close to the most stringent
lower limits to date, coming from LEP - {\it cf.}~\Tabref{lsg-limits}.
Run II at the TEVATRON can improve substantially the results~\cite{Cheung:2000},
especially by accumulating events at high mass.

The H1 collaboration has presented limits~\cite{h1-00}
from deep inelastic scattering at HERA:
\Be
%{\rm H1}: M_s^+ > 0.56\;\TeV \ \ M_s^- > 0.83\;\TeV. %?!
{\rm H1}: M_s^+ > 0.43\;\TeV \ \ M_s^- > 0.64\;\TeV.
\Ee
These limits are somewhat lower than those from LEP or the TEVATRON,
and the next run is expected to improve the sensitivity.

\begin{table}[!htb]
\renewcommand{\arraystretch}{1.20}
\caption{Limits on low scale gravity at 95 \% CL.}
\label{tab:lsg-limits}
  \begin{center}
    \begin{tabular}{|l||c|c|}
       \hline
       Process            &\multicolumn{2}{ c|}{LSG scale $M_s$ (TeV)} \\
                          &  $\ \,\lambda = -1\ \,$ & $\lambda = +1$ \\
       \hline
\multicolumn{3}{|c|}{{D.Bourilkov, JHEP 08 (1999) 006; hep-ph/9907380}} \\
       \hline
$\ee\ra\ee$ (LEP combined $<$ 189 GeV)    &  0.96     &  1.26   \\
       \hline
\multicolumn{3}{|c|}{{S.Mele, E.Sanchez, PRD 61(2000)117901; hep-ph/9909294}} \\
       \hline
$\ee\ra bosons$ (LEP combined $<$ 189 GeV)&  0.96     &  0.93    \\
       \hline
       \hline
\multicolumn{3}{|c|}{{ Winter Conf. 2001 - preliminary}} \\
       \hline
       $\ee\ra\ee$ (L3)                   &  0.98     &  1.06    \\
       \hline
       $\ee\ra\ee$ (OPAL)                 &  1.15     &  1.00    \\
       \hline
       \hline
\multicolumn{3}{|c|}{{ Winter Conf. 2001 - LEP combined preliminary (DB)}} \\
       \hline
       $\ee\ra\ee$                        &  1.28     &  1.13    \\
       \hline
       $\ee\ra\GG$                        &  1.14     &  0.95    \\
       \hline
$\ee\ra\ee$ and $\ee\ra\GG$               &  1.39     &  1.13    \\
       \hline
    \end{tabular}
  \end{center}
%\vspace{-0.2cm}
% $^*$ Combined by DB 
\end{table}

The L3 collaboration has updated the limit on the TeV strings
scale, as shown in~\Tabref{tev-strings}. In this table also the combined
limit extracted from the latest results for the $\GG$ final state
is shown. The limits are still below the result from the first
combined analysis~\cite{Bourilkov:2000}.

\begin{table}[!htb]
\renewcommand{\arraystretch}{1.20}
\caption{Limits on TeV strings at 95 \% CL.}
\label{tab:tev-strings}
   \begin{center}
    \begin{tabular}{|l||c|}
       \hline
       Process        &\multicolumn{1}{ c|}{TeV strings scale $M_S$ (TeV)} \\
       \hline
\multicolumn{2}{|c|}{{ D.Bourilkov, PRD 62(2000)076005; hep-ph/0002172}} \\
       \hline
 $\ee\ra\ee$ (LEP combined $<$ 189 GeV)& 0.63 \\
       \hline
       \hline
\multicolumn{2}{|c|}{{ Winter Conf. 2001 - preliminary}} \\
       \hline
       $\ee\ra\ee$ (L3)                & 0.57  \\
       \hline
  $\ee\ra\GG$ (LEP combined - DB)      & 0.42  \\
       \hline
    \end{tabular}
  \end{center}
%\vspace{-0.2cm}
% {\Large $^*$ Combined by DB} 
\end{table}

\newpage
The last model investigated here is contact interaction coming from
two sets of D-branes~\cite{AntoniadisDBR}.
In the case of muons and taus we can use directly the limits
obtained by the LEP2 $f\bar f$ group for the VV model with positive
interference. The scales are connected as:
\Be
\ee\ra\MM,\TT\; :\ \eta_{\LL} = \eta_{\RR}= \eta_{\LR} = \eta_{\RL} = 1
\Ee
\Be
\L_{VV}^+ \simeq \sqrt{\frac{4\pi}{0.59 g_S}}\cdot M_S.
\Ee
The value of the coupling constant $g_S$ is not known exactly, and
the results are given for $g_S = g_{YM}^2\sim \frac{1}{2}$, with
$g_{YM}$ the gauge coupling, and for the extreme choice
$\frac{g_S}{4\pi} = \frac{1}{128}$.

For electrons the D-brane model predicts a novel type of helicity structure:
\Be
\ee\ra\ee\; :\ 0.75\eta_{\LL} = 0.75\eta_{\RR}\simeq \eta_{\LR} = \eta_{\RL} = 1
\Ee
so we can not use directly limits provided by the individual experiments.
A dedicated analysis is performed, using as input the preliminary measurements
of $\sigma_{tot}$ and $\AFB$ for Bhabha scattering from the four LEP
collaborations. The results are summarized in~\Tabref{dbranes}. They show
a significant improvement compared to the limits in~\cite{AntoniadisDBR},
due to the inclusion of the electron channel and the 2000 data.

\begin{table}[!htb]
\renewcommand{\arraystretch}{1.20}
\caption{Limits on contact interactions in D-brane models at 95 \% CL.}
\label{tab:dbranes}
  \begin{center} 
    \begin{tabular}{|l||c|c|}
       \hline
       Process         &\multicolumn{2}{ c|}{D-brane string scale $M_S$ (TeV)} \\
                       &$g_S = g_{YM}^2\sim \frac{1}{2}$&$\frac{g_S}{4\pi} = \frac{1}{128}$ \\
       \hline
\multicolumn{3}{|c|}{{ Winter Conf. 2001 - LEP combined preliminary (DB)}} \\
       \hline
$\ee\ra\MM$ and $\ee\ra\TT$ &   3.9  &  1.7 \\
       \hline
 $\ee\ra\ee$                &   3.5  &  1.5 \\
       \hline
    \end{tabular}
  \end{center}
%\vspace{-0.2cm}
% $^*$ Combined by DB 
\end{table}

%\newpage
\section{Outlook}

\hspace*{0.7cm}
A short summary:

\begin{center}
\begin{itemize}
  \item[$\clubsuit$] the Standard Model has no difficulty to describe  
        the full set of precise LEP2 measurements  from  130-209 GeV  
        (results at 192--209 GeV are preliminary)
  \item[$\diamondsuit$] the extra dimensions remain hidden  so far  ... \\
        in spite of sensitive searches! 
  \item[$\heartsuit$] many new limits (some deep in the TeV scale!) for physics  \\
        beyond the Standard Model 
  \item[$\spadesuit$] let us hope that some of the future speakers will be more
        lucky and will announce \\
\vspace{0.5cm}
 $$\mathcal{LHC\ NEWS:\ STRING\ DISCOVERY ?}$$
\end{itemize}

%\resizebox{0.85\textwidth}{7.0cm}{\includegraphics{str-ff.eps}}
\end{center}
%
%%%%%%%%%%%%%%%%%%%%%%%%%%%%%%%%%%%%%%%%%%%%%%%%%%%%%%%%%%%%%%%%%%%%%%%%%%%%%%%
% Acknowledgements
%%%%%%%%%%%%%%%%%%%%%%%%%%%%%%%%%%%%%%%%%%%%%%%%%%%%%%%%%%%%%%%%%%%%%%%%%%%%%%%
%
\section*{Acknowledgements}
The author is grateful to the four LEP collaborations for providing their latest
results. Special thanks go to the members of the LEP2 $f\bar f$ group for the
constructive work being done together.
I would like to thank I.~Antoniadis, K.~Benakli, G.~Giudice, S.~Mele,
M.~Peskin and F.~Schrempp for valuable discussions.

%
%%%%%%%%%%%%%%%%%%%%%%%%%%%%%%%%%%%%%%%%%%%%%%%%%%%%%%%%%%%%%%%%%%%%%%%%%%%%%%%
% Bibliography
%%%%%%%%%%%%%%%%%%%%%%%%%%%%%%%%%%%%%%%%%%%%%%%%%%%%%%%%%%%%%%%%%%%%%%%%%%%%%%
%
% Style file to use with mcite.
% Use l3style with just cite.

%\newpage
\bibliographystyle{l3stylem}
\bibliography{%
/afs/cern.ch/l3/paper/biblio/l3pubs,%
/afs/cern.ch/l3/paper/biblio/aleph,%
/afs/cern.ch/l3/paper/biblio/delphi,%
/afs/cern.ch/l3/paper/biblio/opal,%
/afs/cern.ch/l3/paper/biblio/markii,%
/afs/cern.ch/l3/paper/biblio/otherstuff,%
aosta-proc}

\end{document}